\documentclass[useAMS,usenatbib]{mn2e}

\usepackage{natbib}

\usepackage{latexsym,graphicx}%,natbib}
\usepackage{color}
\usepackage{fixltx2e}
\usepackage{verbatim} \usepackage{float}
\usepackage{amsmath,amssymb}
\usepackage{times}

\usepackage{setspace}
\usepackage{rotating}
\usepackage{pdflscape}
\usepackage{subfig}

\usepackage[squaren, Gray, cdot]{SIunits}

\newcommand\hii{H\,{\sc ii} \,}

\newcommand{\mpo}{\textcolor{black}}

\def\apgt{\ {\raise-.5ex\hbox{$\buildrel>\over\sim$}}\ }
\def\aplt{\ {\raise-.5ex\hbox{$\buildrel<\over\sim$}}\ }

\let\oldhat\hat
\renewcommand{\hat}[1]{\oldhat{\mathbf{#1}}}

\renewcommand{\degree}{\ensuremath{^\circ}}

\title[Radio supernova remnants of Wolf-Rayet stars]{Non-thermal radio supernova remnants of exiled Wolf-Rayet stars}

\author[D. M.-A.~Meyer et al.]
       {D. M.-A.~Meyer\thanks{E-mail: dmameyer.astro@gmail.com}$^{1}$, M.~Pohl$^{1,2}$, M.~Petrov$^{3}$ and L.~Oskinova$^{1,4}$   \\
       $^{1}$ Universit\" at Potsdam, Institut f\" ur Physik und Astronomie, Karl-Liebknecht-Strasse 24/25, 14476 Potsdam, Germany\\
       $^{2}$ DESY Platanenallee 6, 15738 Zeuthen, Germany \\        
       $^{3}$ Max Planck Computing and Data Facility (MPCDF), Gießenbachstrasse 2, D-85748 Garching, Germany\\        
       $^{4}$ Department of Astronomy, Kazan Federal University, Kremlevskaya Str 18, Kazan, Russia \\ 
       }

\begin{document}

% Date
\date{Received; accepted}

\maketitle

\label{firstpage}

\begin{abstract} 
\textcolor{black}{
A signification fraction of Galactic massive stars ($\ge 8\,\rm M_{\odot}$) 
are ejected from their parent cluster and supersonically sail away through 
the interstellar medium (ISM). 
The winds of these fast-moving stars blow asymmetric bubbles thus creating a 
circumstellar environment in which stars eventually die with a supernova explosion. 
The morphology of the resulting remnant is largely governed 
by the circumstellar medium of the defunct progenitor star. 
In this paper, we present 2D magneto-hydrodynamical simulations investigating the 
effect of the ISM magnetic field on the shape of the supernova remnants of a 
$35\, \mathrm M_{\odot}$ star evolving through a Wolf-Rayet phase and running with 
velocity $20$ and $40\, \rm km\, \rm s^{-1}$, respectively. 
A $7\, \mu \rm G$ ambient magnetic field is sufficient to modify 
the properties of the expanding supernova shock front and in particular to prevent  
the formation of filamentary structures. 
Prior to the supernova explosion, the compressed magnetic field in the circumstellar medium 
stabilises the wind/ISM contact discontinuity in the tail of the wind bubble. 
A consequence is a reduced mixing efficiency of ejecta and wind materials  
in the inner region of the remnant, where the supernova 
shock wave propagates.
Radiative transfer calculations for synchrotron emission reveal that the 
non-thermal radio emission has characteristic features reflecting the 
asymmetry of exiled core-collapse supernova remnants from Wolf-Rayet progenitors. 
Our models are qualitatively consistent with the radio appearance of several 
remnants of high-mass progenitors, namely the bilateral G296.5+10.0 and the 
shell-type remnants CTB109 and Kes~17, respectively. 
}
\end{abstract}


%%%%%%%%%%%%%%%%%%%%%%%%%%%%%%%%%%%%%%%%%%%%%%%%%%%%%%%%%%%%%%%%%%%%%%%%%%%%%%%%%%%%%%%%%%%
%%%%%%%%%%%%%%%%%%%%%%%%%%%%%%%%%%%%%%%%%%%%%%%%%%%%%%%%%%%%%%%%%%%%%%%%%%%%%%%%%%%%%%%%%%%
%%%%%%%%%%%%%%%%%%%%%%%%%%%%%%%%%%%%%%%%%%%%%%%%%%%%%%%%%%%%%%%%%%%%%%%%%%%%%%%%%%%%%%%%%%%

\section{Introduction}
\label{sect:intro}

Massive stars are born with masses $M_{\star}\ge8\, \rm M_{\odot}$. 
Despite their rareness, they are of prime importance in the cycle of matter in 
the interstellar medium (ISM) of our Galaxy~\citep{langer_araa_50_2012}. 
After a relatively long hydrogen-burning main-sequence phase, they experience a series 
of evolutionary phases characterised by abrupt changes in their surface properties  
(radius, effective temperature, mass-loss rate and wind velocity). 
Those evolutionary phases alternate between hot, possibly eruptive phases of dilute supersonic 
winds~\citep{brott_aa_530_2011a} and colder periods of inflated radius \mpo{with a} dense, 
slow stellar wind~\citep{ekstroem_aa_537_2012}. 
The number and characteristics of the post-main-sequence phases are, amongst other, a
function of the initial mass, the intrinsic rotation~\citep{yoon_443_aa_2005}, and the chemical 
composition of the star~\citep{sanyal_aa_597_2017}. 
These various winds produce shells in the circumstellar medium that develop instabilities 
and eventually collide together~\citep{garciasegura_1996_aa_305f,freyer_apj_594_2003,
freyer_apj_638_2006}. \mpo{They chemically enrich} the ISM and drive turbulence in it, on account
of the large amount of momentum and energy the winds deposit in the stellar surroundings. 
Finally, the majority of massive stars end their life as core-collapse supernova, \mpo{whose} shock wave 
expands into their circumstellar medium~\citep{woosley_rvmp_74_2002}, shaped by stellar winds and 
radiation.

Wolf-Rayet stars are an advance evolutionary stage of stars with initial 
mass $\ge 20\, \rm M_{\odot}$, generally following a supergiant phase.
Their stellar winds are fast, dense, and chemically enriched in C, N and O 
elements~\citep{Hamann2006, bestenlehner_aa_570_2014, Sander2012}. 
The interaction of fast Wolf-Rayet wind with slower wind material expelled 
at previous evolutionary stage results in complex stellar surroundings such as ring 
nebulae~\citep{2010MNRAS.405.1047G,stock_mnras_409_2010,meyer_mnras_496_2020} 
or bipolar \mpo{bubbles}~\citep{gvaramadze_mnras_405_2010}. 
\mpo{A fraction of Wolf-Rayet stars 
are located at high Galactic latitude}~\citep{Munoz2017,Toala2018,Moffat1998}. These fast-moving 
Wolf-Rayet stars \mpo{that left their parent stellar clusters and reached} low-density 
regions of the ISM. There, they \mpo{eventually explode as a core-collapse supernova inside the cavity
carved by the stellar wind}~\citep{franco_pasp_103_1991,
rozyczka_mnras_261_1993,dwarkadas_apj_667_2007}.   
Wolf-Rayet-evolving stars are therefore ideal \mpo{progenitor candidates for}  
core-collapse supernova remnants~\citep{katsuda_apj_863_2018}.

Several mechanisms determine the morphology of the supernova remnants 
of massive progenitors. Clumpiness affecting the shock-wave propagation 
can arise from wind-wind interaction as observed in the supernova remnant 
Cas~A~\citep{vanveelen_aa_50_2009}. 
In addition to instabilities directly developing in the supernova explosion 
itself~\citep{janka_arnps_66_2016}, asymmetries in supernova remnants 
may be a direct consequence of interactions between the expanding shock wave and 
an anisotropic circumstellar medium. \mpo{Of prime importance for the shaping of 
supernova remnants is} the peculiar motion of very 
high-mass progenitors moving through the ISM~\citep{meyer_mnras_493_2020}. 
As an example, RWC 86~\citep{broersen_mnras_441_2014,Gvaramadze_nature} or the 
Cygnus Loop~\citep{aschenbach_aa_341_1999,fang_mnras_464_2017} reveal features 
consistent with the typical characteristics of off-center explosions in 
massive stellar wind bubbles, suggesting that they might have been produced 
by a fast-moving progenitor, see also~\citet{toledo_mnras_442_2014}. 
\mpo{All the numerous mechanisms, that induce deviations from sphericity in supernova 
shock waves, can operate in parallel, providing a huge} parameter space governing 
the evolution of core-collapse supernova remnants. \mpo{Explanations of}
their observed morphologies are subject to degeneracies and alternative scenarios. 
Runaway Wolf-Rayet stars constitute therefore the ideal candidates for the production of 
isolated, \emph{asymmetric} core-collapse supernova remnants~\citep{meyer_mnras_450_2015}.

The structure and properties of the ISM \mpo{are also involved} in the shaping of supernova 
remnants~\citep{ferreira_478_aa_2008}. 
The ISM has an intrinsic filamentary, turbulent, and magnetised nature. Its gravito-turbulent 
evolution, powered by the formation of massive pre-stellar cores, stellar wind outflows, and 
supernova feedback enriching the ISM, drives turbulence in it and participates in the 
formation of the next generation of stars. 
Native ISM magnetic field is an important \mpo{player in} the evolution 
of the circumstellar medium around massive stars. As an example, the internal 
physics of bow shock nebulae around runaway 
stars~\citep{Gvaramadze_2013,meyer_2014a,vanmarle_aa_561_2014,meyer_mnras_464_2017},  
as well as the organisation of supernova 
remnants~\citep{orlando_aa_470_2007,ferreira_478_aa_2008,orlando_apj_678_2008,
schneiter_mnras_408_2010,orlando_apj_749_2012} are partially \mpo{determined by} the local ambient
magnetic field. 
Furthermore, the local direction of magnetic field makes \mpo{thermal conduction}
anisotropic~\citep{balsara_mnras_386_2008,meyer_mnras_464_2017} \mpo{and can suppress}
(magneto-)hydrodynamical instabilities~\citep{viallet_473_aa_2007,vanmarle_aa_561_2014}. 
Importantly, it has been shown that the magnetisation of the ISM strongly elongates stellar 
wind bubbles around static massive stars along the direction of the local field 
lines~\citep{vanmarle_584_aa_2015}. 
The question is therefore, how important \mpo{for the shaping 
of the supernova remnants are the effects of the ISM magnetisation as compared
to those of the motion of} runaway Wolf-Rayet progenitors?

In this work, we investigate, by means of numerical magneto-hydrodynamical (MHD) simulations, 
the effects of a background ISM magnetic field on the morphological evolution of supernova 
remnants generated by runaway massive progenitors. 
We adopt the standard two-dimensional axisymmetric approach developed by 
many authors~\citep{comeron_aa_338_1998,mackey_apjlett_751_2012,meyer_obs_2016}. 
It consists in first modeling the pre-supernova circumstellar medium of massive progenitors 
before launching a supernova blastwave in it~\citep{velazquez_apj_649_2006,chiotellis_aa_537_2012,
vanmarle_aa_541_2012,meyer_mnras_450_2015}. 
We examine the remnant morphologies and perform radiative transfer calculations for their 
non-thermal radio synchrotron emission maps. The mixing of supernova ejecta, stellar winds 
and ISM material is also discussed, comparing models with and without 
ISM magnetic field. Last, we discuss these remnants in the context of cosmic-ray 
acceleration before comparing them to Galactic supernova remnants 
from massive progenitors.

Our study is organised as follows. First, we present the numerical methods used 
for the MHD simulations of supernova remnants of $35\, \rm M_{\odot}$ runaway 
massive stars in Section~\ref{sect:method}. 
We describe our results for the dynamical evolution of both the stellar surroundings 
and the supernova remnant, together with predictive non-thermal radio synchrotron 
emission maps of these objects in Section~\ref{sect:results}. 
We analyse therein the effects of the presence of the ISM magnetic field onto the 
remnants evolution. Our results are further discussed in Section~\ref{sect:discussion}, and 
finally, we present our conclusions in Section~\ref{sect:conclusion}.

%%%%%%%%%%%%%%%%%%%%%%%%%%%%%%%%%%%%%%%%%%%%%%%%%%%%%%%%%%%%%%%%%%%%%%%%%%%%%%%%%%%%%%%%%%%
%%%%%%%%%%%%%%%%%%%%%%%%%%%%%%%%%%%%%%%%%%%%%%%%%%%%%%%%%%%%%%%%%%%%%%%%%%%%%%%%%%%%%%%%%%%
%%%%%%%%%%%%%%%%%%%%%%%%%%%%%%%%%%%%%%%%%%%%%%%%%%%%%%%%%%%%%%%%%%%%%%%%%%%%%%%%%%%%%%%%%%%

\section{Numerical simulations}
\label{sect:method}

This section describes the methods used to perform simulations of the 
circumstellar medium of a $35\, \mathrm M_{\odot}$ massive star evolving up to the 
Wolf-Rayet phase and ending its life in a supernova explosion. We simulate the 
stellar surroundings from the zero-age \mpo{phase} of the progenitor to the 
late phase of supernova remnant evolution, \mpo{varying the velocity 
of the star relative to the ISM} and investigating the role of the ISM magnetic field. 
The simulations are used for further radiative-transfer calculations
of non-thermal radio synchrotron emission.

\subsection{Simulation method for the pre-supernova phase}
\label{sect:hydro}

The pre-supernova circumstellar medium around the progenitor star is the wind-blown bubble generated 
by interaction between the stellar wind and the local ISM. We simulate it \mpo{as described} in~\citet{meyer_mnras_493_2020}. 
We first perform 2D cylindrical, axisymmetric, magneto-hydrodynamics numerical models with a 
coordinate system $[z_{\rm min};z_{\rm max}]\times[O;R_{\rm max}]$ which is mapped 
with a uniform grid of spatial resolution 
$R_{\rm max}/N_{\rm R}$. 
The stellar wind of the $35\, \mathrm M_{\odot}$ star is released at the center of the domain 
into a uniformly distributed ISM. A circular wind zone of radius $20$ cells is filled with the 
wind density profiles,  
\begin{equation}
	\rho_{w}(r) = \frac{ \dot{M} }{ 4\pi r^{2} v_{\rm w} },
\label{eq:wind}
\end{equation}
where $\dot{M}$ is the wind mass-loss rate at different evolutionary phases interpolated from a stellar 
evolutionary track, $r$ is the distance to the origin of the domain, $O$, and $v_{\rm w}$ is the velocity 
of the stellar wind~\citep{comeron_aa_338_1998,vanmarle_apj_734_2011, vanmarle_aa_561_2014}. 

\begin{figure}
        \centering
        \begin{minipage}[b]{ 0.49\textwidth} 
                \includegraphics[width=1.0\textwidth]{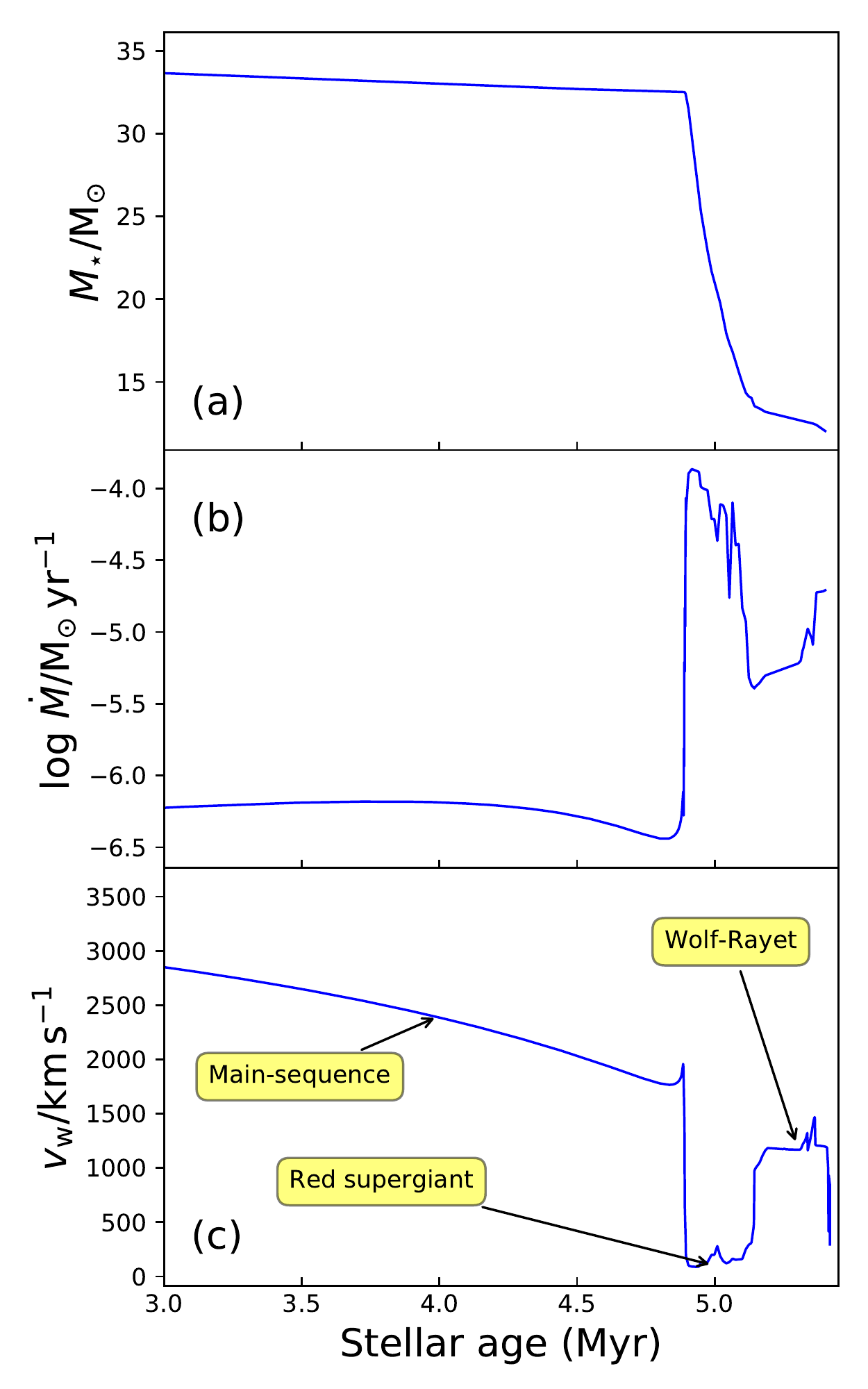}
        \end{minipage}     
        \caption{Stellar properties at the end of the main-sequence and 
    during the post-main-sequence evolution of the $35\, \mathrm M_{\odot}$ star. 
    The panels show the stellar mass (top, panel a), mass-loss rate 
    (middle, panel b), and wind velocity (bottom, panel c) as function 
    of time (in $\rm Myr$).
        }      
        \label{fig:stellar_model}  
\end{figure}

In Fig.~\ref{fig:stellar_model} we show \mpo{the evolutionary path of the 
star and its wind, that we use in the simulations. The stellar mass 
(panel a, in $\rm M_{\odot}$), the mass-loss rate (panel b, in 
$\rm M_{\odot}\, \rm yr^{-1}$), and the terminal wind velocity 
(panel c, in $\rm km\, \rm s^{-1}$) are displayed beginning at the age $3\, \rm Myr$. 
The wind properties of this zero-age-main-sequence, non-rotating 
$35$-$\rm M_{\odot}$ star at Galactic metallicity has been interpolated from 
the Geneva library of stellar models calculated with the {\sc genec} 
code~\citep{ekstroem_aa_537_2012} by means of the online interface 
{\sc syclist}\footnote{https://www.unige.ch/sciences/astro/evolution/en/database/syclist/}. 
The terminal speed, $v_{\rm w}$, is modified for high effective temperatures and massive stars using 
the approximation of~\citet{eldridge_mnras_367_2006},
\begin{equation}    
     v_{\rm w} = \sqrt{ \beta(T) } v_{\rm esc}=\sqrt{ \beta(T)  \frac{ 2GM_{\star}}{R_{\star}}},
\end{equation}
\mpo{where $v_{\rm esc}$ is the escape speed} of the star, $R_{\star}$ the stellar radius, and, 
\begin{equation}
        \beta_{\rm w}(T) =
        \begin{cases}
            1.0   &  \text{if } T \le 10000\, \rm K,  \\
            1.4   &  \text{if } T \le 21000\, \rm K,  \\
            2.65  &  \text{if } T >   21000\, \rm K,  \\ 
        \end{cases}
\end{equation}
a corrective function depending on the temperature $T$.}

\mpo{The star first experiences a rather long main-sequence phase lasting about
$ 4.8\, \rm Myr$, blowing winds with $\dot{M}\approx 10^{-6.2}\, \rm M_{\odot}\, \rm 
yr^{-1}$ and $v_{\rm w}\approx 3000\, \rm km\, \rm s^{-1}$. 
After the long main-sequence phase, \mpo{the star becomes cooler and inflates to become a red supergiant with mass-loss rate} $\dot{M}\approx 10^{-4}\, 
\rm M_{\odot}\, \rm yr^{-1}$ and wind speed $v_{\rm w}\approx 50\, \rm km\, \rm s^{-1}$. It finally 
evolves to the Wolf-Rayet phase, characterised by both a high mass-loss 
rate ($\dot{M}\approx 10^{-5.0}\, \rm M_{\odot}\, \rm yr^{-1}$) and a
large wind speed ($v_{\rm w}\approx 1500\, \rm km\, \rm s^{-1}$). }

\mpo{
To study the circumstellar medium around runaway Wolf-Rayet stars, 
we conducted a series of simulations with varying stellar velocities spanning 
from $v_{\star}=10$ to $v_{\star}=40\, \rm km\, \rm s^{-1}$. The star moves in 
$z$-direction, and we simulate in the frame of the star, ISM gas of number density 
$n_{\rm ISM}\approx0.79\, \rm cm^{-3}$ and temperature $T_{\rm ISM}\approx 8000\, \rm K$ 
as in the \hii regions around hot stars. The ISM material flows in with speed 
$v_{\star}$ at the boundary $z=z_{\rm max}$. Outflow boundaries 
conditions are set at $z=z_{\rm min}$ and $R=R_{\rm max}$, respectively. 
Each value of $v_{\star}$ is explored with and without magnetization of the ISM. 
The ISM magnetic field direction is parallel to the $Oz$ axis, as a direct consequence of the 
simulation geometry, and it is set to $B_\mathrm{ISM}=7\, \mu \rm G$ that is typical value 
for the warm phase of the ISM~\citep{vanmarle_aa_561_2014,vanmarle_584_aa_2015,meyer_mnras_464_2017}.
The flow of material past the stellar wind is characterised by the Alfv\' en speed, 
\begin{equation}
	v_{\rm A} = \sqrt{  \frac{ \bmath{B}_{\rm ISM} \cdot \bmath{B}_{\rm ISM} }{ 4 \pi n m_{\rm H} } },
\label{eq:va}
\end{equation}
that together with the sound speed (Eq.~\ref{eq:cs}) determines the Alfv\' enic and sonic Mach 
number of the stellar wind bubble in the ISM. We list both for each model in Table~\ref{tab:models}. } 

\mpo{A continuity} equation, 
\begin{equation}
	\frac{\partial (\rho Q_{1}) }{\partial t } +  \bmath{ \nabla } \cdot  ( \bmath{v} \rho Q_{1}) = 0,
\label{eq:tracer}
\end{equation}
\mpo{is used to} trace the mixing of stellar wind material into the ISM, with $\rho$ the mass 
density, respectively. Initially, the tracer $Q_{1}$ is set to $Q_{1}(\bmath{r})=1$ in the 
wind and to $Q_{1}(\bmath{r})=0$ in the 
ISM.

\subsection{Supernova explosion}
\label{subsect:sn}

After establishing the circumstellar medium around the pre-supernova  massive star, we simulate  
the supernova explosion as a spherically-symmetric shock wave expanding into the freely-expanding 
stellar wind of the progenitor. The supernova-wind interaction then \mpo{serves} as initial condition 
of a subsequent two-dimensional calculation of the corresponding 
remnant~\citep{meyer_mnras_450_2015,meyer_mnras_493_2020}. 
The properties of the blastwave are parametrised by \mpo{the explosion energy, $E_{\rm ej}=10^{51}\, \rm erg$, and the ejecta mass, 
\begin{equation}
   M_{\rm ej} =  M_{\star} - \int_{t_\mathrm{ZAMS}}^{t_\mathrm{SN}} \dot{M}(t)~ dt - M_{\mathrm{NS}} = 11.64\, \rm M_{\odot},
   \label{eq:co}
\end{equation}
where $t_\mathrm{ZAMS}$ and $t_\mathrm{SN}$ denote the times of zero age and supernova, respectively, and 
$M_\mathrm{NS}=1.4\, \rm M_{\odot}$ is} the mass of the remnant neutron star left behind the supernova 
explosion.  
\textcolor{black}{
Note that we use the canonical explosion energy typically taken in hydrodynamical simulations of 
supernova remnants~\citep{vanveelen_aa_50_2009,vanmarle_mnras_407_2010,vanmarle_aa_541_2012}. However, detailed dedicated studies estimate the energy released throughout the explosion 
of a core-collapse progenitor to be rather in the range $E_{\rm ej}=1$$-$$5\times 10^{50}\, \rm erg$~\citep{smartt_arra_2009,janka_review,moriya_mnras_476_2018}. 
}
A passive scalar, $Q_{2}(\bmath{r})$, obeying the \mpo{continuity} equation, 
\begin{equation}
	\frac{\partial (\rho Q_{2}) }{\partial t } +  \bmath{ \nabla } \cdot  ( \bmath{v} \rho Q_{2}) = 0,
\label{eq:tracers}
\end{equation}
is used to distinguish supernova ejecta from stellar wind or ISM material, by setting 
$Q_{2}(\bmath{r})=1$ in the supernova-ejecta region and $Q_{2}(\bmath{r})=0$ \mpo{otherwise}.

The supernova shock wave is released into the progenitor's stellar 
wind bubble~\citep{whalen_apj_682_2008,zirakashvili_aph_98_2018} by superposing a 1D blastwave 
density profile, $\rho(r)$, onto the pre-supernova wind distribution. 
We used a typical ejecta profile for the early expansion of a core-collapse supernovae. \mpo{It involves a homologuous expansion, $v=r/t$,} the radius of the progenitor 
star's core at the time of the supernova, $r_\mathrm{core}$, and the outermost 
extension $r_\mathrm{max}$ of the blastwave. 
We start the calculations at
\begin{equation}
    t_{\rm max} = \frac{r_{\rm max}}{v_{\rm max}},
\end{equation}
where $v_{\rm max}=30000\, \rm km\, \rm s^{-1}$ is the ejecta velocity at 
$r_\mathrm{max}$~\citep{vanveelen_aa_50_2009}. The value of $r_{\rm max}$ is 
\mpo{determined by the explosion energy and ejecta mass~\citep{whalen_apj_682_2008}. The 
density profile of the ejecta is set as 
\begin{equation}
\rho(r) = \begin{cases}
        \rho_{\rm core}(r) & \text{if $r \le r_{\rm core}$ },               \\
        \rho_{\rm max}(r)  & \text{if $r_{\rm core} < r < r_{\rm max}$},    \\
        \end{cases}
	\label{cases}
\end{equation}
where
\begin{equation}
   \rho_{\rm core}(r) =  \frac{1}{ 4 \pi n } \frac{ (10 E_{\rm ej}^{n-5})^{-3/2}
 }{  (3 M_{\rm ej}^{n-3})^{-5/2}  } \frac{ 1}{t_{\rm max}^{3} },
   \label{sn:density_1}
\end{equation}
is constant, whereas the ejecta density further out} follows a power-law,  
\begin{equation}
   \rho_{\rm max}(r) =  \frac{1}{ 4 \pi n } \frac{ (10 E_{\rm
ej}^{n-5})^{(n-3)/2}  }{  (3 M_{\rm ej}^{n-3})^{(n-5)/2}  } 
\bigg(\frac{r}{t_{\rm max}}\bigg)^{-n},
   \label{sn:density_2}
\end{equation}
with $n=11$~\citep{chevalier_apj_258_1982,truelove_apjs_120_1999}.

\begin{table*}
	\centering
	\caption{
	List of models. The \mpo{columns indicate the velocity of the star, $v_{\star}$, the grid resolution and size in pc, and the sonic and Alf\' enic Mach number of the moving star with respect to the ISM. 
	The runs are labelled "CSM" for the pre-supernova modelling and "SNR" for 
	the remnant simulations, and likewise "HD" for hydrodynamics and "MHD" for magneto-hydrodynamics.   }
	}
	\begin{tabular}{lccccc}
	\hline
	${\rm {Model}}$         &  $v_{\star}$ ($\rm km\, \rm s^{-1}$)  & Grid size                           &    Grid mesh                                    &  $\it M$  &  $\it M_{\rm A}$   \\ 
	\hline   
	Run-35-MHD-20-CSM       &  $20$                                 & $[0;175]\times[-250;100]$           &    $2000 \times 4000\, \mathrm{ cells}$         &  $1.0$  &  $1.16$  \\ 
	Run-35-HD-20-CSM        &  $20$                                 & $[0;175]\times[-250;100]$           &    $2000 \times 4000\, \mathrm{ cells}$         &  $1.0$  &  $1.16$  \\
	Run-35-MHD-40-CSM       &  $40$                                 & $[0;150]\times[-300;100]$           &    $1500 \times 4000\, \mathrm{ cells}$         &  $2.0$  &  $2.32$  \\
	Run-35-HD-40-CSM        &  $40$                                 & $[0;150]\times[-300;100]$           &    $1500 \times 4000\, \mathrm{ cells}$         &  $2.0$  &  $2.32$  \\
	%\hline
	Run-35-MHD-20-SNR       &  $20$                                 & $[0;200]\times[-275;175]$           &    $4000 \times 9000\, \mathrm{ cells}$         &  $1.0$  &  $1.16$  \\
	Run-35-HD-20-SNR        &  $20$                                 & $[0;200]\times[-275;175]$           &    $4000 \times 9000\, \mathrm{ cells}$         &  $1.0$  &  $1.16$  \\
	Run-35-MHD-40-SNR       &  $40$                                 & $[0;200]\times[-330;170]$           &    $~~4000 \times 10000\, \mathrm{ cells}$      &  $2.0$  &  $2.32$  \\
	Run-35-HD-40-SNR        &  $40$                                 & $[0;200]\times[-330;170]$           &    $~~4000 \times 10000\, \mathrm{ cells}$      &  $2.0$  &  $2.32$  \\
	\hline    
%	\hline 
	\end{tabular}
\label{tab:models}
\end{table*}

\mpo{Beyond $r_\mathrm{max}$, the density profile is that of the freely-expanding wind as found in} the pre-supernova wind bubble 
simulations. The ejecta speed at the distance $r_{\rm core}$ from the center of the explosion is~\citep{truelove_apjs_120_1999}
\begin{equation}
   v_{\rm core} = \bigg(  \frac{ 10(n-5)E_{\rm ej} }{ 3(n-3)M_{\rm ej} } \bigg)^{1/2} .
   \label{sn:vcore}
\end{equation}
This 1D ejecta-wind interaction solution is mapped onto the 2D domain \mpo{of the subsequent simulation.} We integrate the equations up to $150\, \rm kyr$ after the supernova.

\subsection{Governing equations}
\label{subsect:goveq}

The dynamics of a magnetised flow is described by the equations 
of ideal magneto-hydrodynamics plus losses and heating by optically-thin radiation,
\begin{equation}
	   \frac{\partial \rho}{\partial t}  + 
	   \bmath{\nabla}  \cdot \big(\rho\bmath{v}) =   0,
\label{eq:mhdeq_1}
\end{equation}
\begin{equation}
	   \frac{\partial \bmath{m} }{\partial t}  + 
           \bmath{\nabla} \cdot \Big( \bmath{m} \textcolor{black}{\otimes} \bmath{v}  + \bmath{B} \textcolor{black}{\otimes} \bmath{B} + \bmath{\hat I}p_{\rm t} \Big) 
            =   \bmath{0},
\label{eq:mhdeq_2}
\end{equation}
\begin{equation}
	  \frac{\partial E }{\partial t}   + 
	  \bmath{\nabla} \cdot \Big( (E+p_{\rm t})\bmath{v}-\bmath{B}(\bmath{v}\cdot\bmath{B}) \Big)  
	  = \Phi(T,\rho),
\label{eq:mhdeq_3}
\end{equation}
and,
\begin{equation}
	  \frac{\partial \bmath{B} }{\partial t}   + 
	  \bmath{\nabla} \cdot \Big( \bmath{v}  \textcolor{black}{\otimes} \bmath{B} - \bmath{B} \textcolor{black}{\otimes} \bmath{v} \Big)  =
	  \bmath{0},
\label{eq:mhdeq_4}
\end{equation}
\mpo{with the linear momentum vector, $\bmath{m}=\rho\bmath{v}$,
and the magnetic-field vector, $\bmath{B}$. }
The total energy of the system reads, 
\begin{equation}
	E = \frac{p}{(\gamma - 1)} + \frac{ \bmath{m} \cdot \bmath{m} }{2\rho} 
	    + \frac{ \bmath{B} \cdot \bmath{B} }{2},
\label{eq:energy}
\end{equation}
where $\gamma=5/3$ is the adiabatic index for ideal gas and $p$ is the thermal pressure. 
The definition of the adiabatic sound speed, 
\begin{equation}
	  c_{\rm s} = \sqrt{ \frac{\gamma p}{\rho} },
\label{eq:cs}
\end{equation}
closes the system, which we integrate using the so-called eight-wave algorithm. 
This second-order unsplit scheme satisfies $\bmath{\nabla} \cdot \bmath{B} = \bmath{0}$.
The time-march of the algorithm obeys the standard Courant-Friedrich-Levy condition that  
is set to $C_{\rm cfl}=0.1$ at the beginning of the simulations. 

The source term,  
\begin{equation}  
	 \itl \Phi(T,\rho)  =  n_{\mathrm{H}}\Gamma(T)   
		   		 -  n^{2}_{\mathrm{H}}\Lambda(T),
\label{eq:dissipation}
\end{equation}
\mpo{accounts for optically-thin radiative cooling, $\itl{\Lambda}(T)$, and heating, $\itl{\Gamma}(T)$.} 
The gas temperature is
\begin{equation}
	T =  \mu \frac{ m_{\mathrm{H}} }{ k_{\rm{B}} } \frac{p}{\rho},
\label{eq:temperature}
\end{equation}
where $\mu=0.61$ is the mean molecular weight, $k_{\rm{B}}$ the Boltzmann constant, and 
$m_{\rm H}$ the proton mass. The hydrogen number density is computed as  
\begin{equation}    
    n_{\mathrm{H}}= \frac{ \rho }{ \mu (1+\chi_{\rm He,Z}) m_{\mathrm{H}} }, 
\end{equation}
with $\chi_{\rm He,Z}$ the mass fraction of all coolants heavier 
than $\rm H$. The functions $\Gamma(T)$ and $\Lambda(T)$ are described in  
details in~\citet{meyer_mnras_464_2017}.

\begin{figure*}
        \centering
        \begin{minipage}[b]{ 0.99\textwidth} 
                \includegraphics[width=1.0\textwidth]{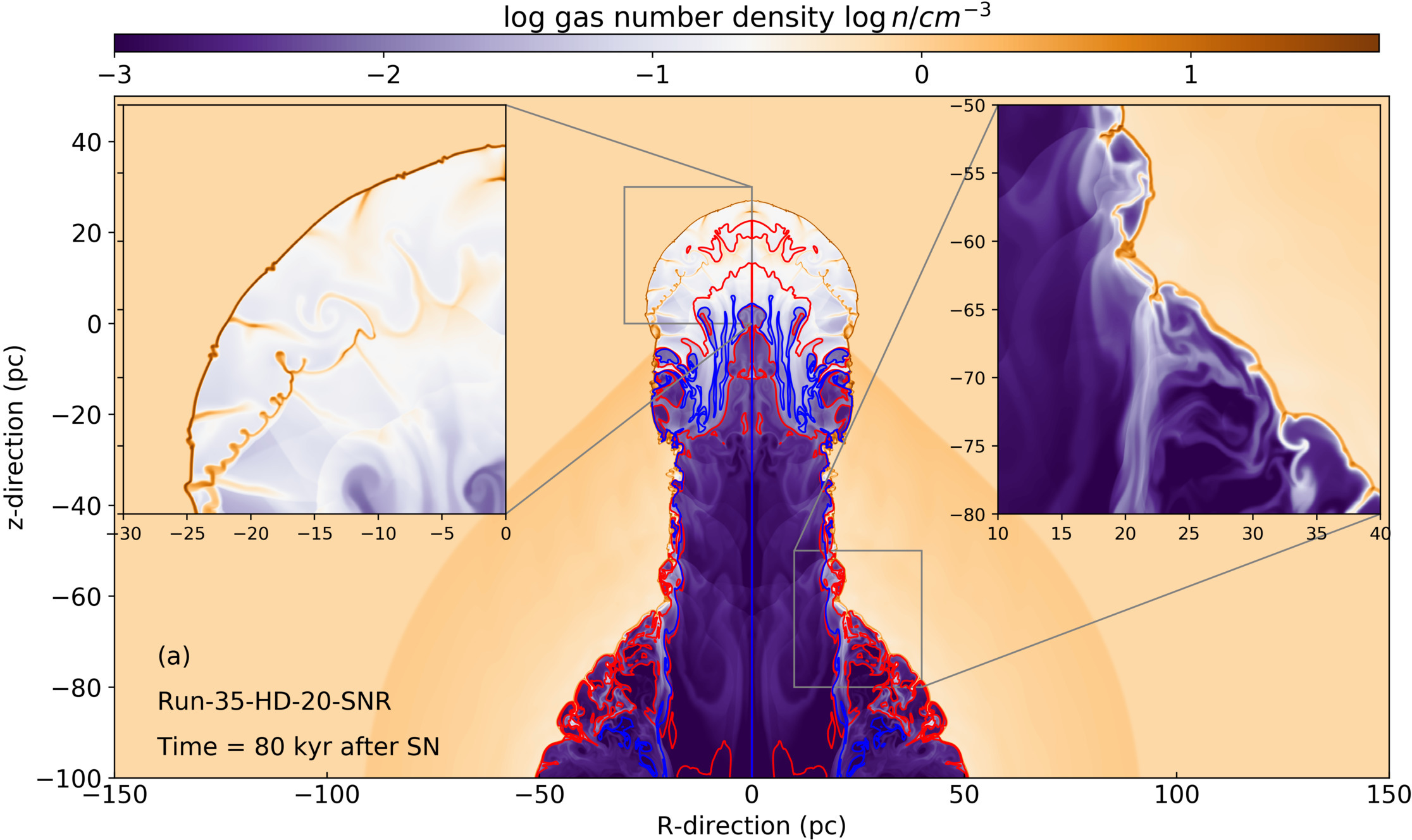}
        \end{minipage}     
        \begin{minipage}[b]{ 0.99\textwidth} 
                \includegraphics[width=1.0\textwidth]{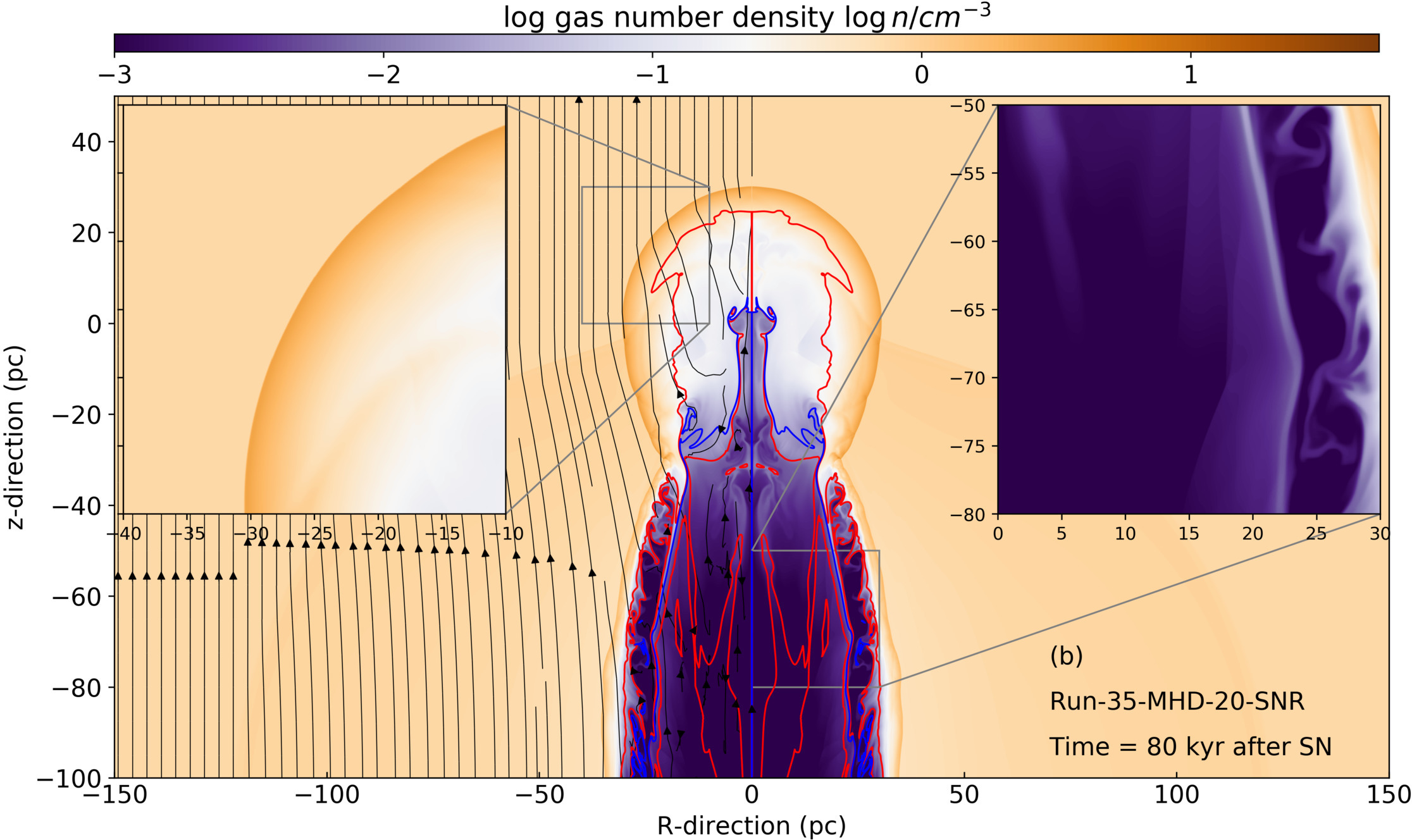}
        \end{minipage}     
        \caption{ 
        Density rendering of the supernova remnant of a $35$-$\mathrm M_{\odot}$ progenitor 
        moving with $v_{\star}=20\, \rm km\, \rm s^{-1}$ through uniform 
        ISM of number density $n_{\rm ISM}=0.78\, \rm cm^{-3}$. 
        Before exploding, the remnant \mpo{passed through} main-sequence, red supergiant and 
        Wolf-Rayet phases~\citep{ekstroem_aa_537_2012}.
        The top panel displays the hydrodynamical model, whereas the bottom one shows the 
        magneto-hydrodynamical \mpo{picture for a $7$-$\mu G$ ambient magnetic field oriented 
        parallel to the stellar motion in z-direction}. 
        Inset boxes highlight the dynamic filamentary structures developing from 
        ejecta-wind-ISM interactions (left inset) and the structure of the stellar 
        wind cavity produced by the progenitor's motion and located behind the center 
        of the explosion (right inset).  
        The red \mpo{lines are iso-temperature contours} ($T =10^{6}$ 
        and $10^{7}\, \rm K$), and the blue contours trace the region with a 
        $10\%$ contribution of supernova ejecta in number density. 
        }      
        \label{fig:zoom_maps_20kms}  
\end{figure*}

%%%%%%%%%%%%%%%%%%%%%%%%%%%%%%%%%%%%%%%%%%%%%%%%%%%%%%%%%%%%%%%%%%%%%%%%%%%%%%%%%%%%%%%%%%%
%%%%%%%%%%%%%%%%%%%%%%%%%%%%%%%%%%%%%%%%%%%%%%%%%%%%%%%%%%%%%%%%%%%%%%%%%%%%%%%%%%%%%%%%%%%
%%%%%%%%%%%%%%%%%%%%%%%%%%%%%%%%%%%%%%%%%%%%%%%%%%%%%%%%%%%%%%%%%%%%%%%%%%%%%%%%%%%%%%%%%%%

\begin{figure*}
        \centering
        \begin{minipage}[b]{ 0.99\textwidth} 
                \includegraphics[width=1.0\textwidth]{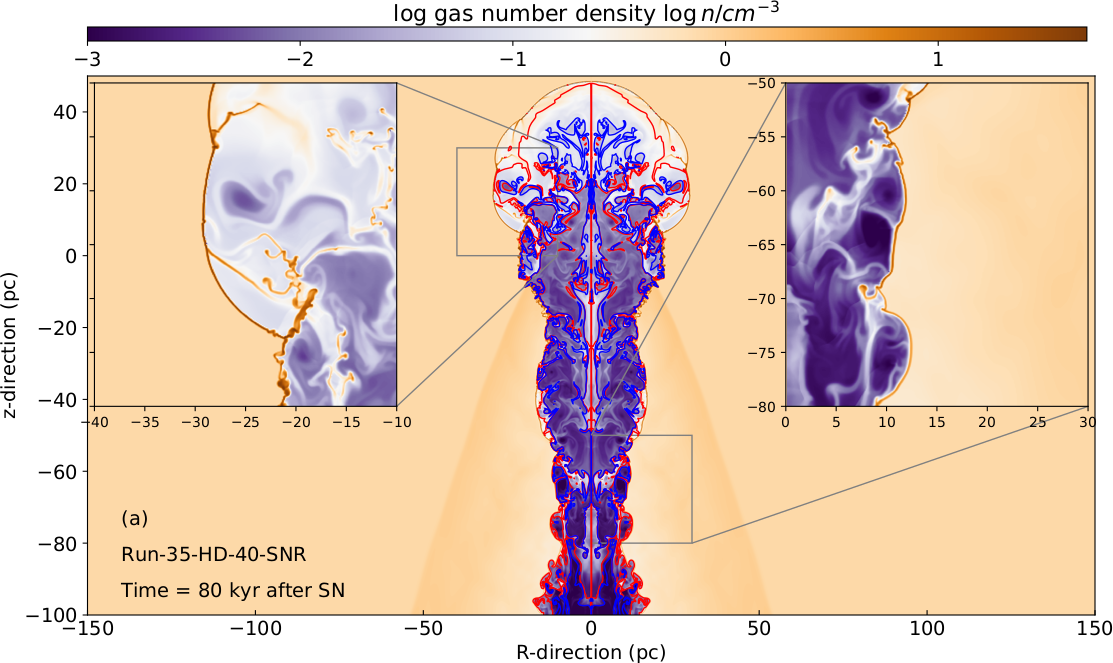}
        \end{minipage}     
        \begin{minipage}[b]{ 0.99\textwidth} 
                \includegraphics[width=1.0\textwidth]{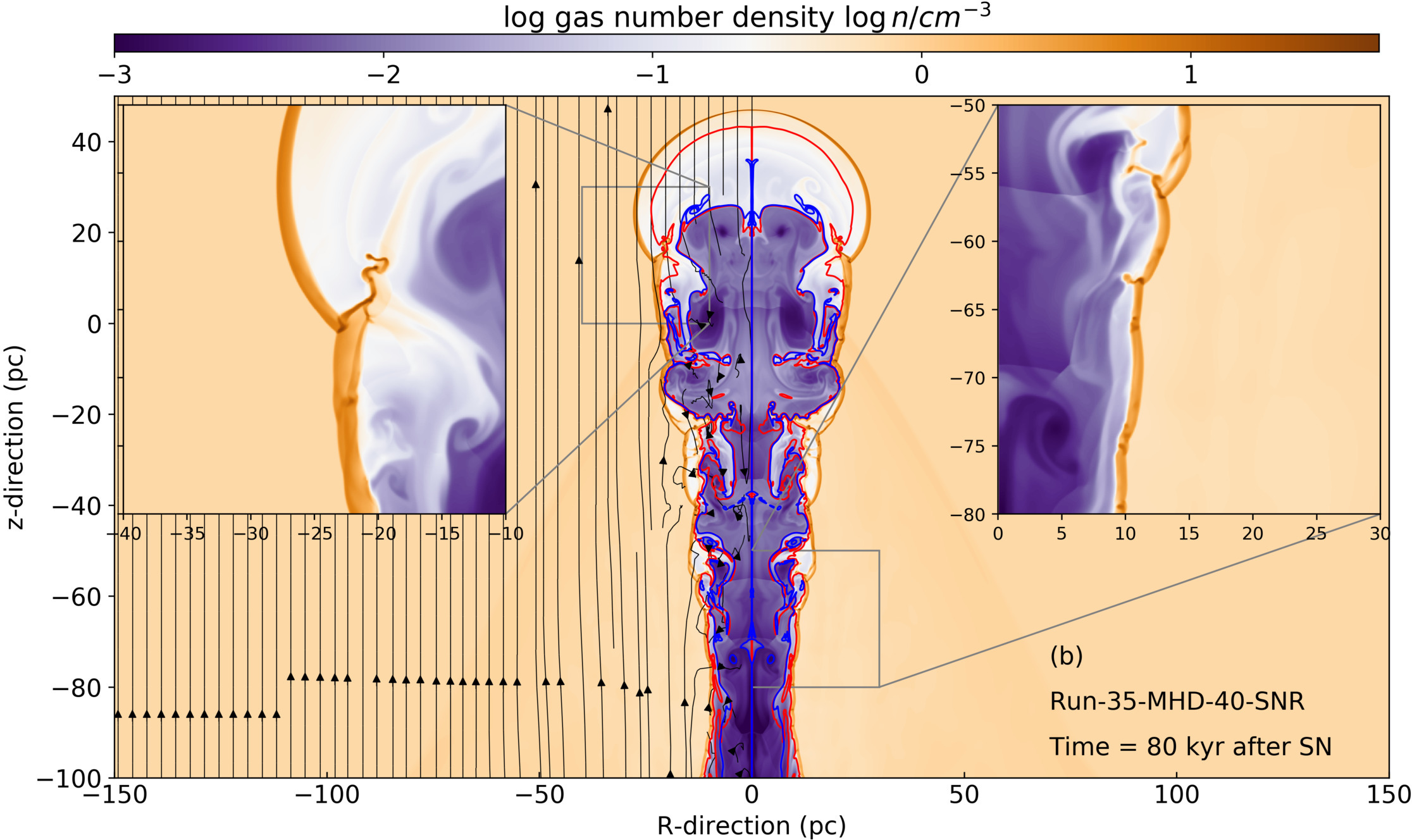}
        \end{minipage}     
        \caption{ 
    As Fig.~\ref{fig:zoom_maps_20kms}, \mpo{but for the supernova remnant of a progenitor moving with} $40\, \rm km\, \rm s^{-1}$.   
        }      
        \label{fig:zoom_maps_40kms}  
\end{figure*}

\subsection{Non-thermal synchrotron radio emission}
\label{sect:radio}

The forward shock of the supernova remnant in particular will accelerate charged particles, such as 
electrons, to high energy. In the presence of magnetic field relativistic electrons produce synchrotron 
emission that is an excellent diagnostic \citep{2008ARA&A..46...89R}.
To permit a comparison with the vast observational data of non-thermal radio synchrotron emission 
from supernova remnants, we produce synthetic emission maps on the basis of our 
magneto-hydrodynamical simulations. 

Energy losses of GeV-scale electrons are likely negligible, and so we assume the electron spectrum,  
\begin{equation}
        N(E) = K E^{-s},
        \label{eq:N}  
\end{equation}
where $E$ denotes the electron energy and the index, $s=2$, is expected for a strong shock. Diffusive 
transport is typically slower than advection in the GeV band, and so the accelerated electron density 
follows the gas density and is in fact proportional to it, if the injection efficiency at the forward 
shock is a constant ~\citep{drury_rpph_46_1983,drury_ssrv_36_1983}.   
\textcolor{black}{
Amongst the several prescriptions for the non-thermal synchrotron emission coefficient of a 
magnetised gas available in the literature, we choose to make use of that 
of~\citet{jun_apj_472_1996}. We refer the reader interested in details regarding 
to our choice of emission coefficient within this core-collapse supernova remnant problem 
in our Appendix~\ref{sync}. 
Therefore, at a given frequency, $\nu$, the radio synchrotron emission coefficient reads, 
\begin{equation}
        j_{\rm sync}(\nu) \propto K^{2-s} p^{s-1} B_{\perp}^{ (s+1)/2 } \nu^{ -(s-1)/2 },
        \label{eq:coeff_bis}  
\end{equation}
which reduces to
\begin{equation}
        j_{\rm sync}(\nu) \propto n^{2-s} p^{s-1} B_{\perp}^{ (s+1)/2 } \nu^{ -(s-1)/2 },
        \label{eq:coeff}  
\end{equation}
where $p$ is the gas thermal pressure and $B_{\perp}$ is the magnetic-field 
component perpendicular to the line of sight. 
}

Let $\vec{l}$ be the unit vector of the observer's line of sight. Defining the viewing 
angle of the observer as $\theta_{\rm obs} =  \angle (\vec{l},\vec{B})$, the total strength 
of the magnetic field and its perpendicular component are obtained as  
\begin{equation}
        B_{\perp} = |\vec{B}| \sin(  \theta_{\rm obs}  )
        \label{eq:BBB}  
\end{equation}
and
\begin{equation}
        |\vec{B}| = \sqrt{ B_{\rm R}^{2} + B_{\rm z}^{2} }.
        \label{eq:BB}  
\end{equation}
\textcolor{black}{
Then, at a given frequency the emission coefficient finally reads,
\begin{equation}
        j_{\rm sync}(\theta_{\rm obs}) \propto n^{2-s} p^{s-1} \Bigg(   |\vec{B}| \sqrt{ 1 - \Big( \frac{ \vec{B}\cdot\vec{l} }{ |\vec{B}| } \Big)^{2}  }  \Bigg)^{ (s+1)/2 },
        \label{eq:final}  
\end{equation}
which we use in our radiative transfer calculations. 
}

For each simulation, we selected snapshots that are representative of the phases of the supernova remnant  
evolution, namely at times $6$, $20$ and $40\, \rm kyr$ after the explosion, respectively. The corresponding 
density, temperature, and magnetic-field distributions are first translated from the two-dimensional 
cylindrical coordinates system to a three-dimensional spherical coordinate system ($r$, $\theta$, $\phi$) with $512^3$ cells and
the same origin, \mpo{for which we rotate the cylindrical solution around the symmetry axis. On each grid zone we pre-calculate the local component of the 
magnetic field that is normal to line-of-sight of the observer. 
Finally, we use a modified version of the radiative transfer code 
{\sc RADMC-3D}\footnote{https://www.ita.uni-heidelberg.de/$\sim$dullemond/software/radmc-3d/} 
to perform ray-tracing integration of the radio synchrotron emission coefficient along a given line-of-sight with aspect angle $\theta_{\rm obs}$. 
The non-thermal radio intensity, 
\begin{equation}
        I = \int_{\rm SNR} j_{\rm sync}(\theta_{\rm obs})  dl,
        \label{eq:intensity}  
\end{equation}
is then used to synthesize} normalised emission maps.

\begin{figure*}
        \begin{minipage}[b]{ 0.48\textwidth} 
                        \includegraphics[width=1.0\textwidth]{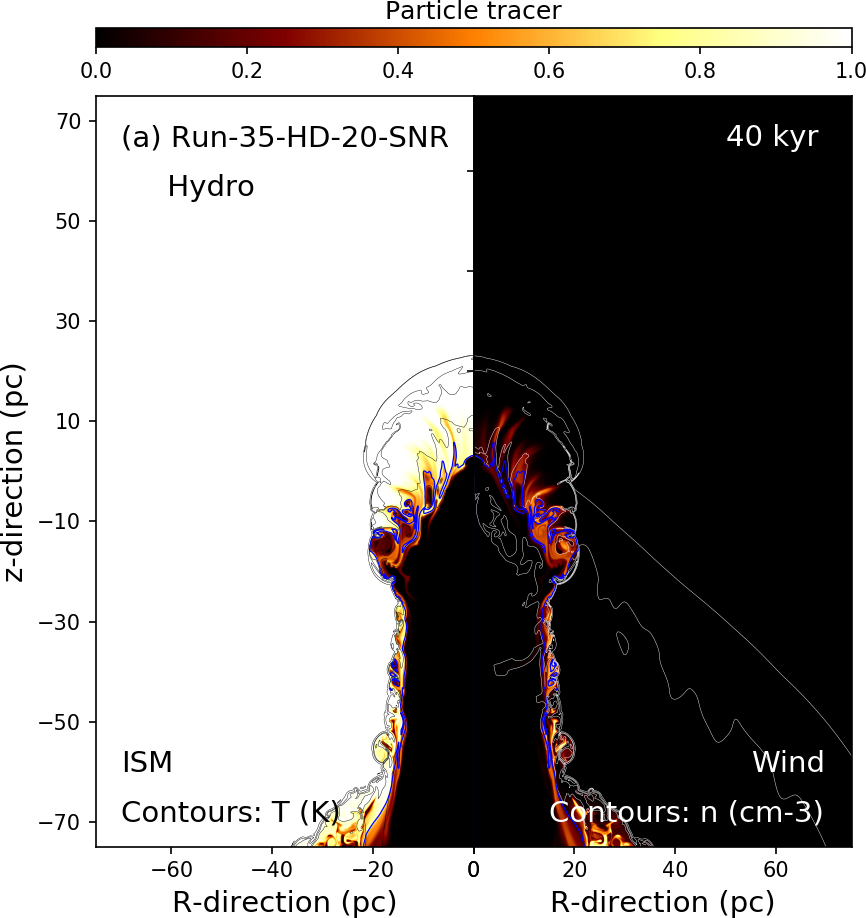}        
        \end{minipage}    
        \begin{minipage}[b]{ 0.48\textwidth} 
                        \includegraphics[width=1.0\textwidth]{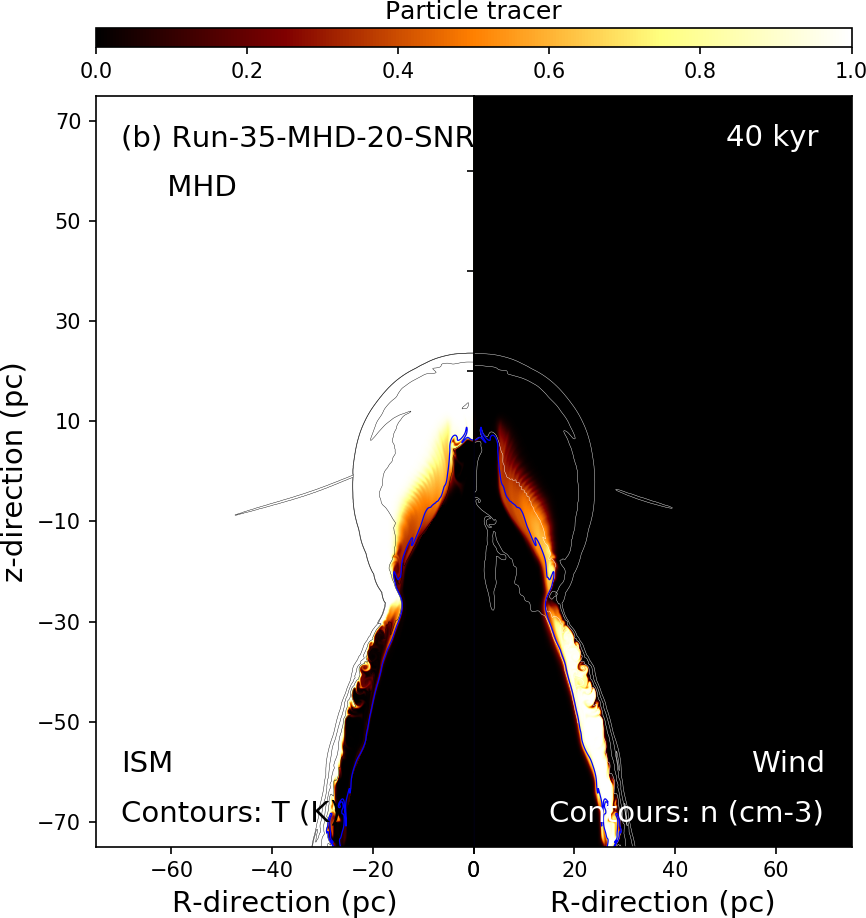}        
        \end{minipage}     \\
        \begin{minipage}[b]{ 0.48\textwidth} 
                        \includegraphics[width=1.0\textwidth]{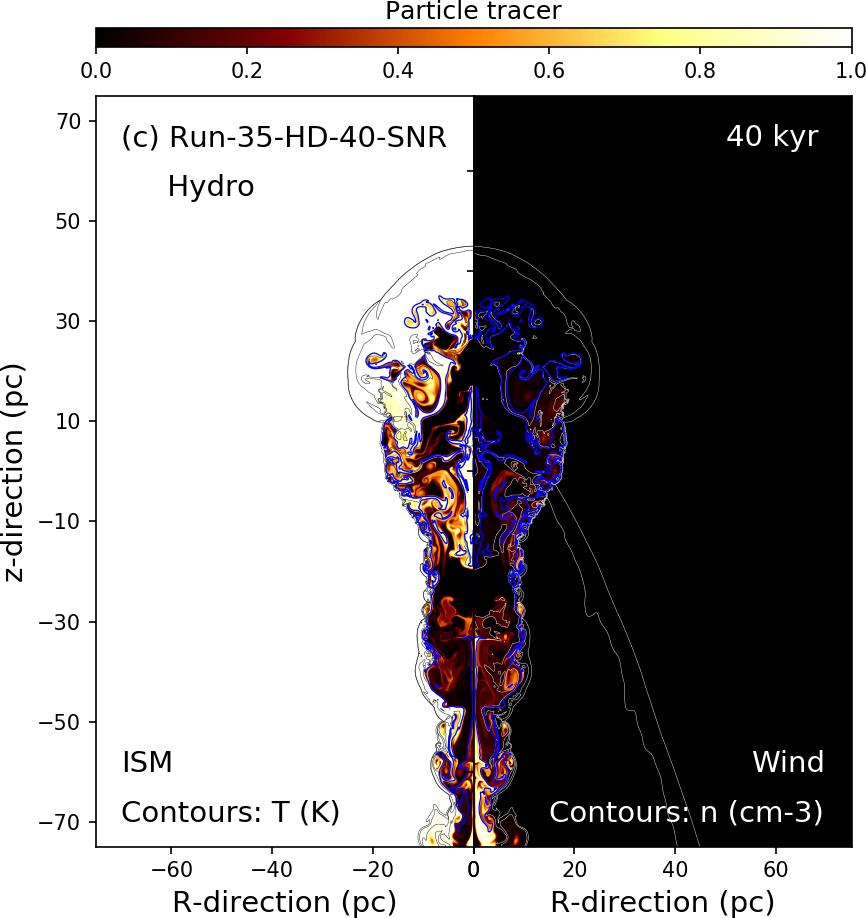}        
        \end{minipage}     
        \begin{minipage}[b]{ 0.48\textwidth} 
                        \includegraphics[width=1.0\textwidth]{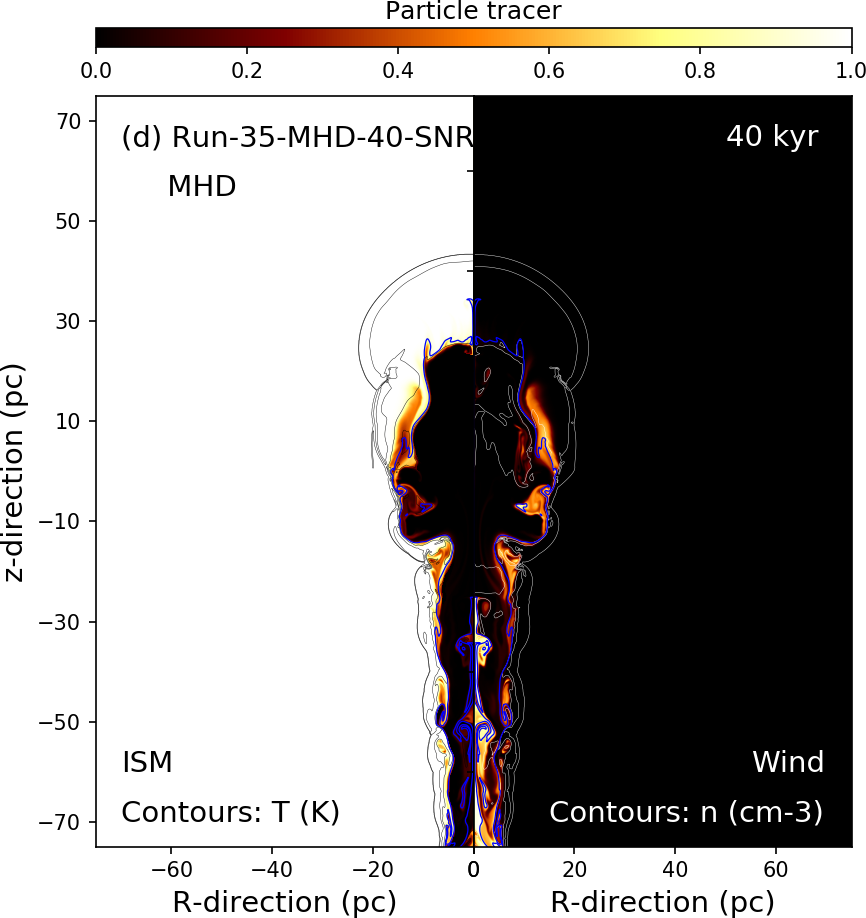}        
        \end{minipage}            
        \caption{ 
    Mixing of material in supernova remnants from massive runaway progenitors. 
    The top row \mpo{displays results for $v_{\star}=20\, \rm km\, \rm s^{-1}$ and the bottom panels are for $v_{\star}=40\, \rm km\, \rm s^{-1}$.
    The left column shows purely hydrodynamical models while the right column is for MHD runs.} Each remnant is shown at time $40\, \rm kyr$ after the supernova explosion. 
    \mpo{Each image gives the ISM fraction ($1-Q_\mathrm{1}$) and the fraction of stellar-wind material ($Q_\mathrm{1}$) in the left and right part, respectively. 
    The black temperature contours have the levels $T = 10^{5}$, $10^{6}$ , $10^{7}\, \rm K$ and
    the 
    white number density contours stand for $n = 1.0$, $10^{1}$, $10^{2}$, $10^{3}\, \rm cm^{-3}$. 
    The blue contours make the locations with $10\, \%$ ejecta fraction by number.}
        }      
        \label{fig:fields_old_snr}  
\end{figure*}

%%%%%%%%%%%%%%%%%%%%%%%%%%%%%%%%%%%%%%%%%%%%%%%%%%%%%%%%%%%%%%%%%%%%%%%%%%%%%%%%%%%%%%%%%%%
%%%%%%%%%%%%%%%%%%%%%%%%%%%%%%%%%%%%%%%%%%%%%%%%%%%%%%%%%%%%%%%%%%%%%%%%%%%%%%%%%%%%%%%%%%%
%%%%%%%%%%%%%%%%%%%%%%%%%%%%%%%%%%%%%%%%%%%%%%%%%%%%%%%%%%%%%%%%%%%%%%%%%%%%%%%%%%%%%%%%%%%

\section{Results}
\label{sect:results}

We present in this section the results for the pre- and post-supernova circumstellar medium 
of a runaway Galactic $35\, \mathrm M_{\odot}$ progenitor star, 
investigate how the stellar motion and the magnetisation of the ISM affect the mixing of 
materials, and present the evolution of their \mpo{projected radio synchrotron emission. }

\subsection{Pre-supernova circumstellar medium of runaway stars}
\label{sect:presn_csm}

Fig.~\ref{fig:zoom_maps_20kms} presents our results for a star \mpo{moving with 
$v_{\star}=20\, \rm km\, \rm s^{-1}$, comparing the hydrodynamical picture
(top) with the magneto-hydrodynamical description} (bottom). 
The red isocontours \mpo{trace $Q_{1}=0.5$, which marks the 
places in the remnant with a 50/50} proportion of ISM and stellar wind. 
The wind of the massive star generated an ovoid bubble of size 
$\sim 100\, \rm pc$ (Fig.~\ref{fig:zoom_maps_20kms}a) \mpo{in which the star is off-centered
on account} of stellar motion~\citep[cf.][]{weaver_apj_218_1977,meyer_mnras_493_2020}.  
The large-scale stellar-wind bow shock is organised according to the classical picture 
of~\citet{weaver_apj_218_1977}, made of an inner termination shock, a contact discontinuity,
and an outer forward shock. 
They distinguish the expanding stellar wind, the hot low-density shocked wind, the cold dense 
ISM gas, and the ambient medium, respectively. The post-main-sequence wind, i.e. the 
red supergiant and Wolf-Rayet materials, are released inside expanding stellar wind and 
develop instabilities at the interface separating the cold and hot gas (Fig.~\ref{fig:zoom_maps_20kms}). 

For a larger speed of the star, $v_{\star}=40\, \rm km\, \rm s^{-1}$, the spherical symmetry is 
broken, and the star reaches the forward shock of its own wind bubble, itself distorted under 
the effects of stellar motion (Fig.~\ref{fig:zoom_maps_40kms}a,b). A chimney of unperturbed 
stellar wind is carved into the layer of shocked ISM, and the post-main-sequence wind is 
blown through the tube (red isocontours). This phenomenon is even more pronounced in the case 
of a very fast progenitor star with $v_{\star}=70\, \rm km\, \rm s^{-1}$~\citep{meyer_mnras_450_2015}. 
Interestingly, once the star has left its wind bubble, direct wind-ISM interaction 
resumes at the distance, 
\begin{equation}
	R_{\rm SO} = \sqrt{ \frac{ \dot{M} v_{\mathrm{w}} }{ 4 \pi n_{\mathrm{ISM}} v_{\star}^{2} } }, 
\label{eq:Ro}
\end{equation}
and a new bow shock forms~\citep{baranov_sphd_15_1971}. An analytic 
estimate for the \mpo{location of the} contact discontinuity of bow shocks reads
\begin{equation}
 \frac{	R({\theta}) }{ R_{\rm SO} } = \frac{ \sqrt{ 3(1-\theta )\mathrm{cotan}( \theta ) } }{ \sin(\theta) },
\label{eq:wilkin}
\end{equation}
where $\theta$ is the angle \mpo{to} the direction of stellar motion~\citep{wilkin_459_apj_1996}. 
This bow shock is in its turn subject to instabilities~\citep{brighenti_mnras_273_1995,brighenti_mnras_277_1995} 
and constitutes the location in which the supernova explosion takes 
place~\citep{brighenti_mnras_270_1994,chiotellis_aa_537_2012,meyer_mnras_450_2015}.

\begin{figure}
        \centering
        \begin{minipage}[b]{ 0.45\textwidth} 
                \includegraphics[width=1.0\textwidth]{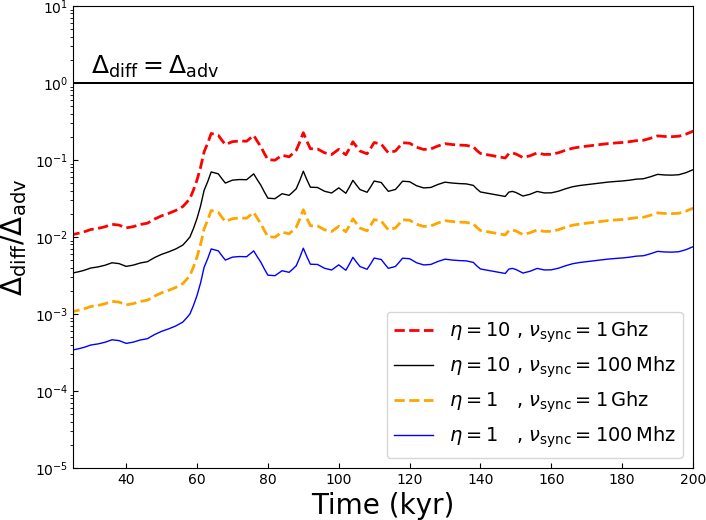}
        \end{minipage}     
        \caption{ 
        \textcolor{black}{
        Comparison between the diffusion timescale, $\Delta_{\rm diff}$, and the advection 
        timescale, $\Delta_{\rm adv}$, of a particle accelerated at the forward shock of the 
        supernova remnant in our simulation Run-35-MHD-40-SNR of a $35$-$\rm M_{\odot}$ 
        progenitor moving with $v_{\star}=40\, \rm km\, \rm s^{-1}$.   
        The quantity $\Delta_{\rm diff}/\Delta_{\rm adv}$ is plotted as a function of \mpo{remnant age}
        for several values of $\eta$ and 
        synchrotron frequency $\nu_{\rm sync}$, respectively. 
        \mpo{Values well below unity imply} that diffusion can be neglected in the estimate of 
        radio emission maps.         }
        }      
        \label{fig:diffusion}  
\end{figure}

The \mpo{interstellar magnetic field has strong impact on the distribution of} shocked ISM, 
as demonstrated in~\citet{vanmarle_584_aa_2015}. As an example, in Figs.~\ref{fig:zoom_maps_20kms}b 
and~\ref{fig:zoom_maps_40kms}b one clearly sees that the layer of shocked ISM is 
puffed up along the local magnetic field as a result of the damping of Alfv\' en waves~\citep{vanmarle_584_aa_2015}. 
\mpo{The effect is weaker for a fast-moving progenitor on account of the lower ratio of magnetic 
and ram pressure} (Fig.~\ref{fig:zoom_maps_40kms}b). 
The \mpo{entire interior structure of the wind bubble is elongated when the star moves quickly. The 
magnetic-field lines are aligned with the termination shock and discontinuities, providing additional 
pressure that modifies the circumstellar gas dynamics (Figs.~\ref{fig:zoom_maps_20kms}b). 
It can also damp instabilities at the contact discontinuity between hot shocked wind and cold 
shocked ISM gas, primarily in the tail of the wind bubble. 
Our models combine the asymmetry in the stellar wind bubbles of moving stars~\citep{meyer_mnras_450_2015} 
with the magnetic-pressure effect that were previously explored for static stars}~\citep{vanmarle_584_aa_2015}.

\subsection{Supernova remnants}
\label{sect:snr}

Fig.~\ref{fig:fields_old_snr} displays \mpo{the structure of} the supernova remnant 
at time $40\, \rm kyr$ after the explosion. Each panel corresponds to a different simulation, 
with density plotted on the left-hand part of the panel and the temperature 
on the right-hand part of the panel, respectively. 
The blue isocontour marks the regions with 10\% abundance of ejecta in number density.
The \mpo{left panels are derived from hydrodynamical simulations, to be compared with MHD results on the right.} 
The stellar velocities are $v_{\star}=20\, \rm km\, \rm s^{-1}$ (top) and $v_{\star}=40\, \rm km\, \rm s^{-1}$ 
(bottom), respectively.

The shape of the supernova remnants are governed by the distribution of the pre-supernova 
circumstellar medium~\citep{meyer_mnras_450_2015,meyer_mnras_493_2020}. 
Figs.~\ref{fig:fields_old_snr}b,d show that the faster the progenitor moves through the 
ISM, the sooner the supernova shock wave interacts with the termination shock of the progenitor's 
wind bubble. 
For smaller $v_{\star}$, the elongated shape of the MHD wind bubble permits the Wolf-Rayet 
wind to expands freely into the unperturbed red supergiant stellar wind and to generate 
by wind-wind collision a ring of dense swept-up material~\citep{meyer_mnras_496_2020}, 
inside of which the blastwave is subsequently released and expands spherically. 
A similar situation has been explored for static progenitor in the context of 
Cas A~\citep{vanveelen_aa_50_2009}. 
This phenomenon is particularly prominent for small ISM density, $n_{\rm ISM}\ll1$, since the radius 
of the main-sequence wind termination shock is much larger~\citep{vanmarle_584_aa_2015}, and 
so is the region filled by the last free-streaming wind.

The effects of the ISM magnetic field are \mpo{also more pronounced for low progenitor speed,}
$v_{\star}=20\, \rm km\, \rm s^{-1}$. 
The supernova remnant shock wave rapidly interacts with the wind bubble in the progenitor's direction of
motion, and it is first reverberated towards the center of the explosion and subsequently
channelled into the wind cavity carved during the main-sequence phase, inducing a hot region 
hosting a lot of mixing of wind, ejecta and shocked ISM gas (Fig.~\ref{fig:fields_old_snr}a). 
The reflections are different in the MHD case, \mpo{where they occur both parallel and normal to the progenitor's 
motion, on account} of the tubular shape of the shocked-wind region 
(Fig.~\ref{fig:fields_old_snr}b,d). 
A different morphology arises for fast-moving progenitors, with a rather 
unmixed lobe of shocked ISM ahead of the stellar motion, and a channelled region of 
mixed ejecta and wind material \mpo{in the tail} (Figs.~\ref{fig:fields_old_snr}c,d).

\begin{figure}
        \centering
        \begin{minipage}[b]{ 0.47\textwidth} 
                \includegraphics[width=1.0\textwidth]{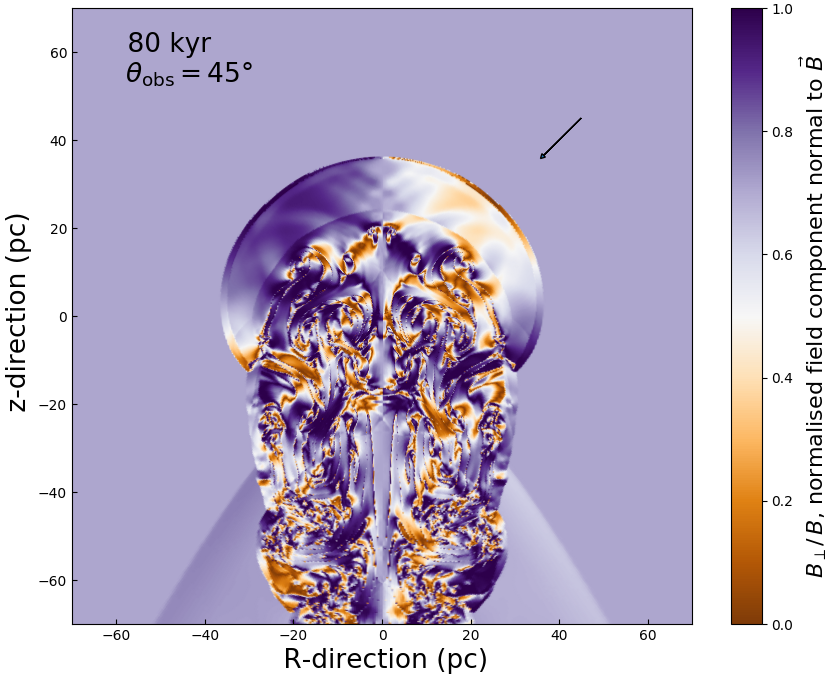}
        \end{minipage}         
        \caption{ 
        Normalised projection of the magnetic field \mpo{perpendicular to the line of sight, $B_{\perp}/B$, for model Run-35-MHD-40 model with progenitor speed
        $v_{\star}=40\, \rm km\, \rm s^{-1}$ at time $80\, \rm kyr$ after 
        the supernova explosion. 
        The black arrow marks the direction of the observer's line-of-sight, making an 
        angle $\theta_{\rm obs}=45\degree$ with the $z$-axis. }
        }      
        \label{fig:radio_40kms}  
\end{figure}

\subsection{Effect of ISM magnetic field on the remnant's properties}
\label{sect:emission}

\mpo{Magnetic field changes the morphology of the pre-supernova stellar 
wind bubble, which will eventually influence the structure} of the remnant and the mixing of material in it. 
\mpo{At a magnetized shock, the component of the magnetic field along the shock normal 
remains unchanged, and that in the shock plane is compressed}~\citep{shu_pavi_book_1992}. This results in an increased 
magnetic pressure in the shocked ISM and, consequently, in an enlargement 
of the bubble \mpo{perpendicular} to the direction of motion, \mpo{which for static wind bubbles has been demonstrated by}~\citet{vanmarle_584_aa_2015}. 
\mpo{Further effects are a reduced compression ratio of the forward shock and a 
puffing-up of the shocked ISM gas layer.}  
Although the MHD jump conditions imply that the \mpo{ISM field is not compressed ahead of the 
star,} since the cylindrical coordinate system imposes a parallel field, we can not 
exclude that this is \mpo{an artefact.} We refer the reader to the thorough discussion 
in~\citet{meyer_mnras_464_2017}. 
Fully three-dimensional simulations of both the pre-supernova and remnant phase of moving massive 
progenitor star are necessary to address this question in appropriate detail.

\begin{figure*}
        \centering
        \begin{minipage}[b]{ 0.9\textwidth} 
                \includegraphics[width=1.0\textwidth]{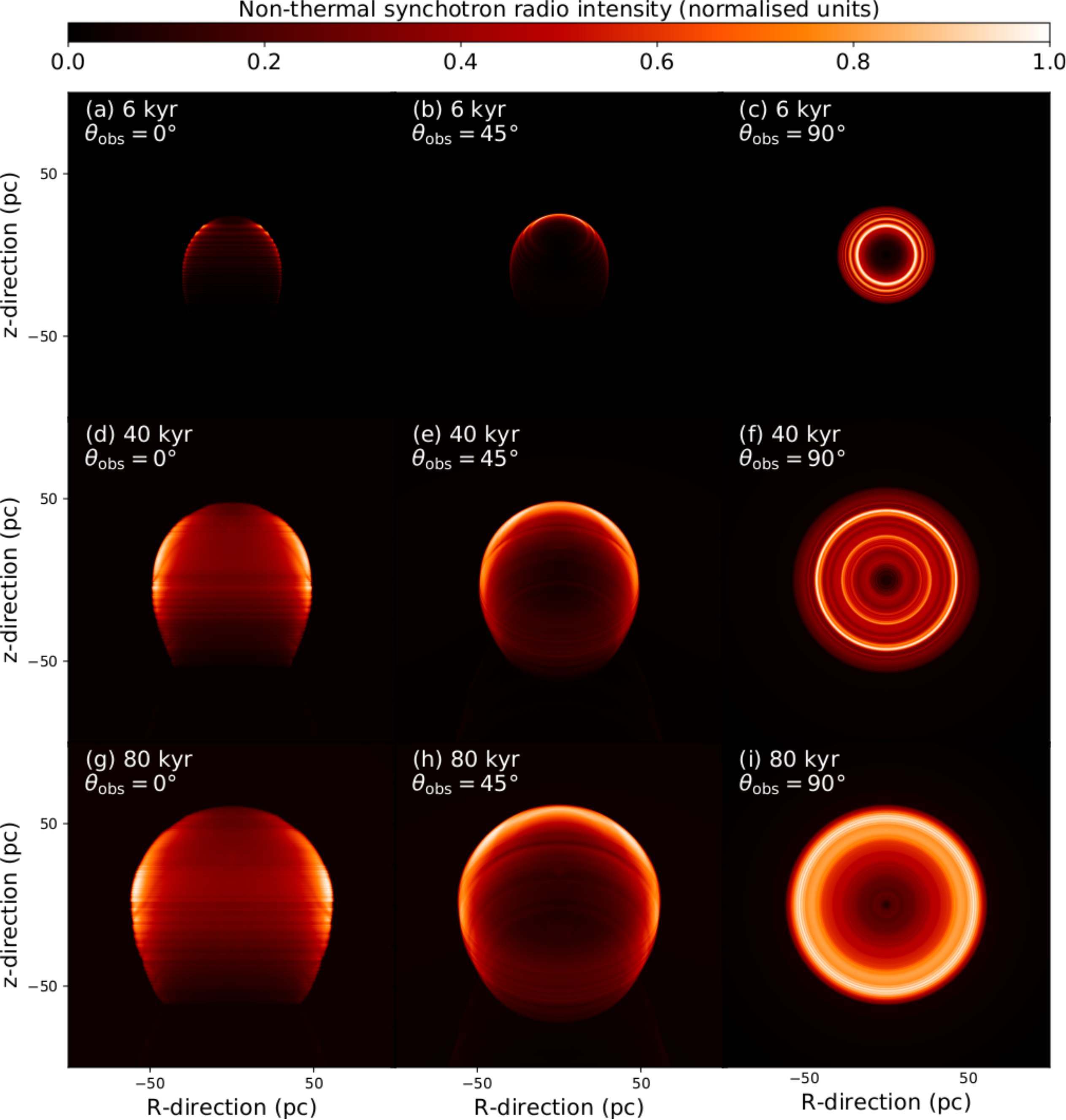}
        \end{minipage}     
        \caption{ 
        \mpo{Normalised maps of radio synchrotron intensity of supernova remnants 
        with progenitor speed $v_{\star}=20\, \rm km\, \rm s^{-1}$ at times $6\, \rm kyr$ (top), 
        $40\, \rm kyr$ (middle )and $80\, \rm kyr$ (bottom) after the supernova explosion, respectively. 
        The viewing angle between the equatorial plane and the 
        line of sight is} $\theta_{\rm obs}=0\degree$ (left), $\theta_{\rm obs}=45\degree$ (middle),
        and $\theta_{\rm obs}=90\degree$ (right). 
        }      
        \label{fig:radio_map_20kms}  
\end{figure*}

\subsection{Radio synchrotron emission maps}
\label{sect:maps}

\subsubsection{\textcolor{black}{Basic consideration}}
\label{sect:maps_nodiff}

\textcolor{black}{
We perform a back-of-the-envelope estimate of the validity of our 
assumptions regarding the radio synchrotron emission. 
We select one-dimensional cross-section of the flow variables $\rho$, $p$ and $T$ 
originating from the center of the explosion and following the direction of 
stellar motion, along the axis of symmetry, $Oz$, in the models, respectively.  
We then measure the time-dependent position, $R_{\rm FS}(t)$, and 
speed, $v_{\rm FS}(t)$, \mpo{of the forward shock, using the shock-finder algorithm of 
the {\sc ratpac} code~\citep{telezhinsky_aph_35_2012,telezhinsky_aa_552_2013,bhatt_apj_aph_2020}.} 
The radiation synthesis is based on the assumption that the forward shock accelerates a 
certain fraction of particles passing through it, and that the subsequent transport of 
radio-emitting electrons in the downstream region is entirely advective. }

\textcolor{black}{
In our radiation transfer model, diffusion is neglected and we shall now verify this assumption. 
Upstream of the forward shock the density of electrons re-accelerated as cosmic-rays exponentially decreases, 
$N(E,x)\propto e^{-x/x_{\rm c}(t)}$ with length scale,
\begin{equation}
	x_{\rm c}(t) = \frac{ D(E) }{ v_{\rm FS}(t) }, 
\label{eq:L2}
\end{equation}
where $D(E)$ the diffusion coefficient and $E$ the energy of 
the electrons. For simplicity, we shall assume that the diffusion coefficient is a multiple of 
the Bohm limit, $D(E)\approx \eta D_{\rm Bohm}(E)= \eta\,c/3\,r_\mathrm{L}(E)$, where $r_\mathrm{L}$ 
denotes the Larmor radius of the electrons and $\eta$ is a scalar controlling the . Using the characteristic 
frequency of synchrotron radiation, Eq.~\ref{eq:L2} can be rewritten as
\begin{equation}
	\frac{x_{\rm c}(t)}{\mathrm{pc}} \simeq \frac{\eta}{3000}\sqrt\frac{\nu_\mathrm{syn}}{10\,\mathrm{GHz}}\left(\frac{v_{\rm FS}(t)}{3000\,\mathrm{km/s}}\right)^{-1} \left(\frac{B(t)}{10\,\mathrm{\mu G}}\right)^{-1.5}.
\label{eq:L2a}
\end{equation}
It is evident that for reasonably well developed cosmic-ray scattering ($\eta\ll 1000$) the 
precursor of radio-emitting electrons is tiny compared to the size of the system and produces 
a negligible contribution of synchrotron emission. Note that a significant abundance of magnetic 
field perpendicular the shock surface will slow down diffusive transport away from the shock and 
hence reduce $\eta$.}

Diffusion in the downstream region may transport electrons beyond the discontinuity to the 
ejecta region. Within time $t$, advective transport displaces electrons from the shock by the 
distance $\Delta_\mathrm{adv} \simeq t\,v_{\rm FS}(t)/4$. Diffusive transport displaces by 
$\Delta_\mathrm{dif} \simeq \sqrt{ D(E) t }$, and so ignorability of diffusion requires  
\begin{equation}
t\,v_{\rm FS}(t)/4 \gg \sqrt{ D(E) t }\quad\Rightarrow\ 
t\,v_{\rm FS}(t)\gg 16 x_{\rm c}(t) .
\label{eq:L1}
\end{equation}
\mpo{With $x_c$ as given in Eq.~\ref{eq:L2a} we find that the condition is likely met.}
Fig.~\ref{fig:diffusion} plots $\Delta_{\rm diff}/\Delta_{\rm adv}$ as a function of time 
for our simulation model Run-35-MHD-40-SNR and several values of $\eta\le10$ and 
$\nu_{\rm sync\le1\, \rm Ghz}$. \mpo{To be noted from the figure is} that 
$\Delta_{\rm diff}/\Delta_{\rm adv} \ll 1$ for times $\le 200\, \rm kyr$, and consequently 
we can ignore cosmic ray diffusion in the production of radio emission maps. 

\begin{figure*}
        \centering
        \begin{minipage}[b]{ 0.9\textwidth} 
                \includegraphics[width=1.0\textwidth]{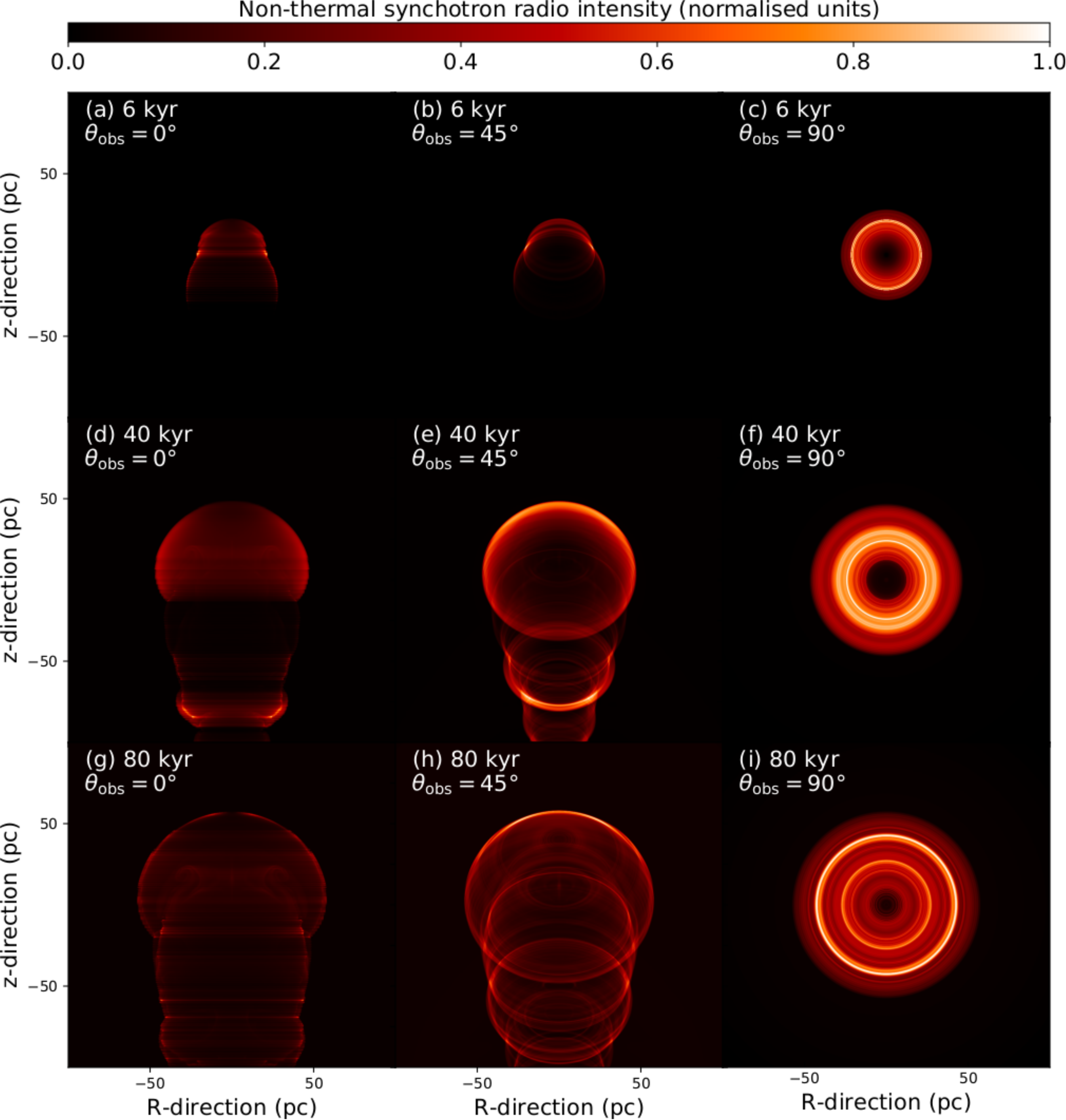}
        \end{minipage}     
        \caption{ 
    As Fig.~\ref{fig:radio_map_20kms} but here for model 
    Run-35-MHD-40-SNR with progenitor moving with $40\, \rm km\, \rm s^{-1}$.   
        }      
        \label{fig:radio_map_40kms}  
\end{figure*}

\subsubsection{\textcolor{black}{Non-thermal emission maps}}
\label{sect:maps_maps}

We generate non-thermal radio emission maps from our MHD models of 
supernova remnants at representative \mpo{points of time in}
the evolution of the supernova remnants. 
\mpo{Using the procedure described in section~\ref{sect:radio} we
pre-compute for each viewing angle, $\theta_{\rm obs}$, the distribution 
of magnetic field normal to the line of sight, $B_{\perp}$. An 
illustrative example based on model Run-35-MHD-40 at time $80\, \rm kyr$ is 
given in Fig.\ref{fig:radio_40kms}. Then we compute the 
radio intensity with} our modified version 
of the {\sc radmc3d} code. 
\mpo{In Fig.~\ref{fig:radio_map_20kms} we show intensity maps for supernova 
remnant ages $6\, \rm kyr$ (top), $20\, \rm kyr$ (middle )and $80\, \rm kyr$ 
(bottom) for a progenitor moving with $v_{\star}=20\, \rm km\, \rm s^{-1}$.
We selected three viewing angles to the equatorial plane,
$\theta_{\rm obs}=0\degree$ (left), $\theta_{\rm obs}=45\degree$ (middle) and 
$\theta_{\rm obs}=90\degree$ (right) and normalised the background-substracted maps. }
Fig.~\ref{fig:radio_map_40kms} displays corresponding radio maps for $v_{\star}=40\, \rm km\, \rm s^{-1}$.

Model Run-35-MHD-20 with progenitor moving at $v_{\star}=20\, \rm km\, \rm s^{-1}$ 
\textcolor{black}{
traces the expanding supernova blastwave that is distorbed by its interaction with the 
circumstellar medium. 
}
Later, $40\, \rm kyr$ after the explosion, the expanding shock wave has reached the 
unperturbed ISM and is fairly bright there. The radio arcs are now larger than at time 
$20\, \rm kyr$ and the brightest region on the sides have a bilateral morphology. 
Note that the maps are background-subtracted, and the region of shocked stellar wind 
\mpo{may be dimmer than the galactic radio background, leaving only the filamentary 
arcs prominently} visible. 
For an inclination angle $\theta_{\rm obs}=45\degree$ the 
remnants look rounder and more bubbly. \mpo{One and the same remnant can appear with bilateral or arced
structures} depending on $\theta_{\rm obs}$. 
For $\theta_{\rm obs}=90\degree$ the observer's line-of-sight is aligned with the direction 
of stellar motion, and the projected remnant appears as a ring-like structure 
in the sky, \mpo{on account of the two-dimensional nature of the 
simulations.} Since the magnetic field is parallel to the stellar motion, 
$B_{\perp}\simeq 0$ in the shocked ISM, and the brightest emission originates from the 
regions of mixing, primarily in the reflected shock wave.

The radio intensity maps of our simulation Run-35-MHD-40 with progenitor 
speed $v_{\star}=40\, \rm km\, \rm s^{-1}$ are shown in Fig.~\ref{fig:radio_map_40kms}. 
The emission are brighther than in model Run-35-MHD-20 as a result of the faster progenitor 
star producing stronger shocks in its supernova remnant. 
At time $6\, \rm kyr$ after the explosion, the shock wave has already been greatly distorted by the 
Wolf-Rayet circumstellar material and has lost sphericity to become as ovoid-like structure. 
Later in time, the shock wave adopts a hour-glass-like shape that appears spherical \mpo{in the radio map for $\theta_{\rm obs}=45\degree$ and to a lesser degree at $\theta_{\rm obs}=0\degree$. 
%
%Similarly, projection effects are at work and make the bilateral shapes produced with a considered viewing angle  rounder at, e.g. $\theta_{\rm obs}=45\degree$. 
%
After $40\, \rm kyr$, the remnant has an bulb-like morphology which arises 
from both the shock wave expansion into the ISM and the channeling of the shock wave into 
the low-density cavity of unshocked stellar wind in the tail. 
The radio intensity peak shifts} to the location where the shock wave intercepts 
the trail of stellar wind (Fig.~\ref{fig:radio_map_40kms}d,e).   
The density in the region of wind-ISM interaction is more important than in the case of a 
progenitor moving with $v_{\star}=20\, \rm km\, \rm s^{-1}$, and so the stabilising 
effect of the magnetic field inside of the remnant is reduced. Hence, more ring-like 
structures appear in the emission maps, for example in Fig.~\ref{fig:radio_map_40kms}h. 
\mpo{We would observe} a series of concentric rings 
for $\theta_{\rm obs}=90\degree$, each of them corresponding to a ring in the trail 
of shocked stellar wind interacting with the channelled supernova shock wave. 
\mpo{There is more variety and complexity in the radio appearance for} faster-moving progenitors.

%%%%%%%%%%%%%%%%%%%%%%%%%%%%%%%%%%%%%%%%%%%%%%%%%%%%%%%%%%%%%%%%%%%%%%%%%%%%%%%%%%%%%%%%%%%
%%%%%%%%%%%%%%%%%%%%%%%%%%%%%%%%%%%%%%%%%%%%%%%%%%%%%%%%%%%%%%%%%%%%%%%%%%%%%%%%%%%%%%%%%%%
%%%%%%%%%%%%%%%%%%%%%%%%%%%%%%%%%%%%%%%%%%%%%%%%%%%%%%%%%%%%%%%%%%%%%%%%%%%%%%%%%%%%%%%%%%%

\section{Discussion}
\label{sect:discussion}

This section discusses the limitations of our method, compares our results to earlier results, and further examines our findings in 
the context of particle acceleration. Finally, we compare our results with observational data.

\subsection{Model limitations}
\label{sect:limitations}

As any numerical study, our method suffers from \mpo{simplifications that} limit the realism of our results. 
\mpo{The most obvious one is the cylindrical 
coordinate system with rotational invariance,} which intrinsically imposes a symmetry axis to the models. This approach is 
convenient in the modelling of the circumstellar medium of massive stars and their subsequent 
supernova remnants~\citep{franco_pasp_103_1991,
rozyczka_274_MNRAS_1995,comeron_aa_326_1997,comeron_aa_338_1998,vanmarle_aa_444_2005,
vanmarle_aa_469_2007,ferreira_478_aa_2008,vanmarle_aa_561_2014,2019A&A...625A...4G}, \mpo{at the expense of forcing an directional alignment of the motion of the progenitor, the local ISM magnetic field, and progenitor's axis of rotation. Only fully three-dimensional simulations  
permit flexibility in the directional arrangement~\citep[e.g.][]{katushkina_MNRAS_465_2017,katushkina_MNRAS_473_2018}, but they are far too expensive to permit} scanning the parameter space 
of remnants from massive progenitors. 

%The list of physical processes which have been taken into account in our simulations isnot exhaustive. 
Supernova remnants from massive progenitors are multi-phase regions composed of a warm, magnetized interstellar medium through which the progenitor star moves, the evolving stellar wind, and a hot component produced by the interaction between the supernova shock wave and
the material of the wind bubble and the shocked ISM gas. Supernova remnants \mpo{may be located} close 
to dense, cold molecular \mpo{clouds that can further affect their evolution and modify the gas chemistry. } 
%This ideally would require a treatment of the chemical reactions happening therein using a detailed network of reactions such as {\sc krome}. 
%
The pressure of the cosmic rays accelerated in the supernova 
remnant~\citep{ferrand_apj_789_2014}, anisotropic heat 
transfer~\citep{orlando_aa_444_2005}, photoionizing progenitor radiation, or 
the turbulence in the ISM should also be included in the 
models~\citep{moranchel_mnras_472_2017,villagran_mnras_491_2020}, \mpo{but that 
is far beyond the scope of the current study and may be 
considered in future work.} %s using the Lagrangian Particles module of the {\sc pluto} code~\citep{vaidya_apj_865_2018}. 
\textcolor{black}{
Last, note that intrinsic dense molecular~\citep{zhou_apj_743_2011,zhou_apj_791_2014,
zhou_apj_831_2016} or low-density 
components~\citep{arias_aa_622_2019,arias_aa_627_2019} of the 
ambient medium are an additional, in some context an even preponderant 
element to take into account in the shaping of core-collapse supernova remnants. 
}

\subsection{Comparison with previous works}
\label{sect:previous_works}

This study \mpo{extends earlier work beginning with~\citet{meyer_mnras_450_2015} on 
hydrodynamical models of supernova remnants of runaway stars with initial $10$, $20$ 
and $40\, \rm M_{\odot}$,} that end their lives as red supergiants and generate 
Cygnus-Loop-like nebulae. 
The second study of this series explored the appearance of wind nebulae and remnants of 
a $60$ $\rm M_{\odot}$ progenitor star going through Luminous-Blue-Variable and Wolf-Rayet
phases, with emphasis on the mixing of material inside the 
remnant~\citet{meyer_mnras_493_2020}. What is new and different in the present study is 
the inclusion of the ISM magnetic field during both the pre- and the post-supernova phase, 
together with the post-processing \mpo{of radio synchrotron intensity maps.} 
The effect of ISM magnetisation on the environment of massive stars 
has been investigated by~\citet{vanmarle_584_aa_2015}, albeit without distinguishing 
ejecta from wind and ISM gas as we do by means of passive scalar tracers. Moreover, our 
models include a state-of-art stellar evolutionary model for the wind history of the 
$35$ $\rm M_{\odot}$ star that we concentrate on.

\subsection{Comparison with observations}
\label{sect:obs}

\subsubsection{\textcolor{black}{Remnants of core-collapse supernovae }}

\citet{katsuda_apj_863_2018} determined several properties, such as the distance 
and the progenitor mass, of core-collapse supernova remnants in the Milky Way 
and in low-metallicity dwarf galaxies such as the Large and Small Magellanic Clouds. 
They found that most of the identified remnants of massive progenitors in the Galaxy
have a zero-age main-sequence mass $\ge 22.5\, \rm M_{\odot}$. There is a general agreement 
between predictive stellar evolution models that the progenitor exploded with such mass, 
either as a red supergiant or as a Wolf-Rayet star, although more exotic situations such as 
blue supergiant progenitor star exist~\citep{Hillebrandt_nature_327_1987}. 
Note that, in the context of massive binary systems, the explosion of the component  
can kick the companion, producing runaway stars~\citep{2021arXiv210105771L}. 
Our models explore the Wolf-Rayet possibility using a zero-age main-sequence 
$35\, \rm M_{\odot}$ star.  
As we concentrate on the evolution of rather older remnants, about $6$$-$$80\, \rm kyr$ 
after the explosion, our predictions are applicable to the objects listed between 
Kes~79 and W51C in Table~1 of~\citet{katsuda_apj_863_2018}.

Our models constitute baseline models to be further tailored to specific supernova 
remnants, in particular Kes 79, G350.1-0.3, G292.0+1.8, 
RX J1713.7-3946, Kes 79, G290.1-0.8, 3C391, W44, G284.3-1.8 or CTB109. 
\textcolor{black}{
Note that C, N and O enriched material, witness of post-main-sequence winds 
from massive stars, has been directly observed in the
remnant G296.1-0.5~\citep{castro_apj_734_2011}, making it an evident 
candidate of a supernova remnant with Wolf-Rayet progenitor that is worth 
exploring numerically with simulations like ours. 
}
As underlined by~\citet{katsuda_apj_863_2018}, the distribution of core-collapse 
supernovae in the Galaxy does not fit any initial mass function, 
which suggest that there should be many more \mpo{unidentified remnants 
our simulations would be} applicable to. 
Note also that models for core-collapse remnants do not generally apply to 
the so-called historical supernova remnants since these are mostly of
type~Ia~\citep{green_lnp_598_2003}, 
except for Cas~A~\citep{vanveelen_aa_50_2009,zhou_apj_865_2018} and 
RWC~86~\citep{Gvaramadze_nature}, respectively.

The space motion of the progenitor is the other fundamental ingredient 
\mpo{ of our simulations,} together with the zero-age main-sequence mass.  
If the massive progenitor is at rest, then 
its circumstellar wind bubble remains spherical~\citep{weaver_apj_218_1977}. The star, and 
the center of its subsequent explosion are located at its 
center~\citep{freyer_apj_638_2006,dwarkadas_apj_667_2007}. A sub-sonic motion of the 
progenitor will off-center the remnant with respect to the wind bubble without changing its 
overall appearance~\citep{meyer_mnras_493_2020}. 
Hence, remnants from slowly-moving progenitor star should reflect the spherical symmetry of their 
circumstellar medium. However, as in-situ star formation does not seem to be an obvious 
route to explain isolated Wolf-Rayet stars~\citep{gvaramadze_424_mnras_2012,meyer_mnras_496_2020}, 
static massive stars should live and die inside of their parent star-formation 
region, where they would participate in the regulation of \mpo{subsequent star} formation~\citep{paron_aa_498_2009}. 
The feedback from stellar winds and/or jets of other (young) 
stellar objects~\citep{bally_prpl_2007,fendt_apj_692_2009,fendt_apj_774_2013} will affect the 
medium in which massive stars form~\citep{murray_mnras_475_2018}, evolve and die. 
This should result in a very complex morphology, possibly further complicated by 
ISM cavities and enhanced levels of turbulence,
in the ambient medium hosting a huge mix of material in 
a super bubble in which the supernova subsequently 
explodes~\citep{vanmarle_aa_541_2012}.

\subsubsection{\textcolor{black}{Comparison with specific objects}}

Most supernova remnants in the Milky Way lie within $5\degree$ of the galactic plane. 
The dilute ISM at high galactic latitudes makes (circumstellar) shocks weaker, resulting 
in fainter supernova remnants such as, for example, the rather evolved radio source 
G181.1+9.5~\citep{kothes_aa_597_2017}.  This latitude-dependence of the radio surface brightness 
of supernova remnants is known as the $\Sigma$-D relation~\citep{caswell_mnras_187_1979}. 
The \mpo{modelled $35\, \rm M_{\odot}$ runaway star moving with 
$v_{\star}=40\, \rm km\, \rm s^{-1}$ travels about $ 220\, \rm pc$ before exploding, 
about the same as} the height of G181.1+9.5 above the galactic plane ($\approx 250\, \rm pc$). 
Dedicated simulations would be highly desirable to explore the differences \mpo{in the 
radio properties between supernova remnants of runaway progenitors in the Galactic 
place and those at} higher galactic latitudes, as well as the effects of metallicities. 
\textcolor{black}{
Our non-thermal radio emission maps authorise a couple of further, direct comparisons with 
supernova remnants of massive, evolved progenitors, namely G296.5+10.0, the shell-type 
remnants CTB 109, and Kes~17.}

\textcolor{black}{
First, G296.5+10.0 is a supernova remnant of core-collapse origin, \mpo{confirmed by} the trace of 
magnetised wind in which the supernova shock wave expands, and the 
presence of a neutron star therein~\citep{harvey_apj_712_2010}. Radio observations 
with the Australia Telescope Compact Arrary at $1.4\,$GHz reveal a bipolar shape, 
that is qualitatively similar to \mpo{those in} our model with velocity $v_{\star}=20\, \rm km\, \rm s^{-1}$, 
Run-35-MHD-20-SNR, at time $6\, \rm kyr$, when the shock wave interacts with the stellar-wind 
bow shock and accelerates electrons (Fig.~\ref{fig:radio_map_20kms}a). 
\textcolor{black}{
Similarly, Fig.~\ref{fig:radio_map_40kms}a also resembles greatly the $7\, \rm kyr$ old bilateral 
supernova remnant G296.5+10.0, implying that its progenitor might have been a rather fast 
moving star of mass $20$$-$$40\, \rm M_{\odot}$.  
}
Secondly, the shell-type remnant CTB~109 is the remnant of a core-collapse supernova remnant 
of a $30$-$40\rm M_{\odot}$ progenitor, which matches the mass range of the $35$-$\rm M_{\odot}$ 
stellar model used in this study. Its age is around
$14000\, \rm yr$, see~\citet{katsuda_apj_863_2018}. 
}

\textcolor{black}{
\mpo{According to our results,} CTB~109 should be surrounded by a circumstellar structure, i.e. a red-supergiant wind 
bubble engulfing a Wolf-Rayet ring, with which the supernova shock 
wave interacts, although its overall shape has been reproduced in the context of a type~Ia explosion, 
i.e. without the presence of a dense circumstellar wind bubble generated by a massive  
progenitor~\citep{bolte_aa_582_2015}. Its opened shell appearance, e.g. as seen with 
the Canadian Galactic Plane Survey at $1420\, \rm MHz$~\citep{kothes_apj_746_2012}, 
is similar to our model Run-35-MHD-SNR at times $6$-$40\, \rm kyr$ 
(Fig.~\ref{fig:radio_map_40kms}b,e). Note also that CTB~109 is a
\mpo{hadronic gamma-ray emitter~\citep{castro_apj_756_2012}. 
The last example is} the supernova remnant Kes~17 that is less than $40000\, \rm yr$ old and with a progenitor mass 
$25$-$30\rm M_{\odot}$ consistent with the Wolf-Rayet scenario~\citep{katsuda_apj_863_2018}. 
Its double-arced morphology observed with the Australian Telescope Array at $20\, \rm cm$ 
resembles our model Run-35-MHD-40-SNR at times $40\, \rm kyr$ (Fig.~\ref{fig:radio_map_40kms}c). 
}

\section{Conclusion}
\label{sect:conclusion}

\mpo{
We explore the formation, structure, and radio signatures of supernova remnants 
of massive, Wolf-Rayet-evolving supernova progenitors ejected from their parent cluster 
and moving through the interstellar medium (ISM) of the Milky Way. Our study concentrates 
on the coupled impact of stellar motion and the magnetisation of the ISM. 
We perform magneto-hydrodynamical simulations over the entire stellar lifetime, as they 
successively evolve through a long main-sequence phase, a red supergiant and a Wolf-Rayet 
phase, and eventually spawn a core-collapse supernova remnant.  
Numerical models are performed with the {\sc pluto} code~\citep{mignone_apj_170_2007,
migmone_apjs_198_2012} by simulating the circumstellar medium of massive stars, into which 
we launch a core-collapse supernova shock wave. We follow its interaction with the stellar 
surroundings and the local ambient ISM supported by an organised $7\, \mu \rm G$ 
magnetic field. 
Considering two speeds of the runaway progenitors and running the simulations up to the 
oldest evolutionary phase of the supernova remnants, $150\, \rm kyr$ after the explosion, 
their morphologies, their structures, the mixing of material happening in them are explored.  
}

%Although the overall morphology of the remnants of runaway Wolf-Rayet progenitors moving with 
%$20$ or $40\,\rm km\, \rm s^{-1}$ remains similar with or without the presence
%
The presence of an ISM magnetic field profoundly affects the gas properties. 
\textcolor{black}{ 
Prior to the supernova explosion, the compressed magnetic field in the circumstellar medium 
stabilises the wind/ISM contact discontinuity in the tail of the bubble. 
Indeed, compressed magnetic field in the outer remnant stabilises and elongates the wind/ISM contact 
discontinuity of the cavity of unshocked stellar wind, in which the ejecta is channelled. 
A consequence is a reduced mixing efficiency of ejecta and evolved stellar-wind material 
enriched in C, N and O elements in the inner region of the remnant, where the supernova shock 
wave propagates. 
Moreover, after the supernova explosion, the density downstream of the supernova shock front is 
reduced in our MHD simulations when it propagates into the pristine ambient medium, on account of 
the damping of turbulence.  
This must influence the acceleration processes of cosmic-ray electrons and protons in 
supernova remnants from massive progenitors and will be investigated in future 
works~\citep{bhatt_apj_aph_2020}. 
We emphasize the need for a careful treatment of the gas microphysics to 
properly simulate young supernova remnants interacting with circumstellar structures.  
This particularly applies to runaway massive progenitors whose supernova shock front 
is reverberated towards the center of the explosion, generating a complex region made 
of shocks, discontinuities, and filamentary structures, in which non-thermal 
particles can be accelerated.
}

Last, using our modified version of the radiative transfer code 
{\sc radmc/3d}~\citep{dullemond_2012} we produced synthetic radio-intensity 
maps \textcolor{black}{showing} projected arcs and filaments that we interpret as a 
morphological characteristic of supernova remnants of fast-moving Wolf-Rayet stars. 
\textcolor{black}{
Our radio predictions are qualitatively in accordance with the morphology of 
several core-collapse remnants, such as the bilateral G296.5+10.0, as well as the 
shell-type supernova remnants CTB~109 and Kes~17, identified as originating 
from $25$-$40\, M_{\odot}$ progenitors~\citep{katsuda_apj_863_2018} which might 
have undergone a Wolf-Rayet phase. 
}
Our simulations and predictions regarding the non-thermal emission of supernova 
remnants from massive progenitors are relevant for and \mpo{may be applied} 
to the various galactic and extragalactic core-collapse remnants~\citep{katsuda_apj_863_2018}.

%%%%%%%%%%%%%%%%%%%%%%%%%%%%%%%%%%%%%%%%%%%%%%%%%%%%%%%%%%%%%%%%%%%%%%%%%%%%%%%%%%%%%%%%%%%
%%%%%%%%%%%%%%%%%%%%%%%%%%%%%%%%%%%%%%%%%%%%%%%%%%%%%%%%%%%%%%%%%%%%%%%%%%%%%%%%%%%%%%%%%%%
%%%%%%%%%%%%%%%%%%%%%%%%%%%%%%%%%%%%%%%%%%%%%%%%%%%%%%%%%%%%%%%%%%%%%%%%%%%%%%%%%%%%%%%%%%%

\section*{Acknowledgements}

\textcolor{black}{
Authors are grateful to the referee, P.~Velazquez, for comments on synchrotron emission 
which greatly improved the quality of the manuscript. 
}
The authors thank Allard Jan van Marle from Ulsan National Institute of Science and Technology 
for his kind advices on MHD simulations of the surroundings of massive stars. 
The authors acknowledge the North-German Supercomputing Alliance (HLRN) for providing HPC 
resources that have contributed to the research results reported in this paper. 
M.~Petrov acknowledges the Max Planck Computing and Data Facility (MPCDF) for providing 
data storage resources and HPC resources which contributed to test and optimise the {\sc pluto} code. 
L.M.O. acknowledges partial support by the Russian Government Program of Competitive Growth of Kazan Federal University. 

\section*{Data availability}

This research made use of the {\sc pluto} code developed at the University of Torino  
by A.~Mignone (http://plutocode.ph.unito.it/) and of the {\sc radmc-3d} code developed 
at the University of Heidelberg by 
C.~Dullemond (https://www.ita.uni-heidelberg.de/$\sim$dullemond/software/radmc-3d/).
The figures have been produced using the Matplotlib plotting 
library for the Python programming language (https://matplotlib.org/). 
The data underlying this article will be shared on reasonable request to the corresponding author.

%%%%%%%%%%%%%%%%%%%%%%%%%%%%%%%%%%%%%%%%%%%%%%%%%%%%%%%%%%%%%%%%%%%%%%%%%%%%%%%%%%%%%%%%%%%
%%%%%%%%%%%%%%%%%%%%%%%%%%%%%%%%%%%%%%%%%%%%%%%%%%%%%%%%%%%%%%%%%%%%%%%%%%%%%%%%%%%%%%%%%%%
%%%%%%%%%%%%%%%%%%%%%%%%%%%%%%%%%%%%%%%%%%%%%%%%%%%%%%%%%%%%%%%%%%%%%%%%%%%%%%%%%%%%%%%%%%%

% Bibliography style for the bibtex
\bibliographystyle{mn2e}

\footnotesize{
% Create the reference Section using BibTeX:
\bibliography{grid}
}

%%%%%%%%%%%%%%%%%%%%%%%%%%%%%%%%%%%%%%%%%%%%%%%%%%%%%%%%%%%%%%%%%%%%%%%%%%%%%%%%%%%%%%%%%%%
%%%%%%%%%%%%%%%%%%%%%%%%%%%%%%%%%%%%%%%%%%%%%%%%%%%%%%%%%%%%%%%%%%%%%%%%%%%%%%%%%%%%%%%%%%%
%%%%%%%%%%%%%%%%%%%%%%%%%%%%%%%%%%%%%%%%%%%%%%%%%%%%%%%%%%%%%%%%%%%%%%%%%%%%%%%%%%%%%%%%%%%

\appendix

\begin{figure*}
        \centering
        \begin{minipage}[b]{ 1.0\textwidth} 
                \includegraphics[width=1.0\textwidth]{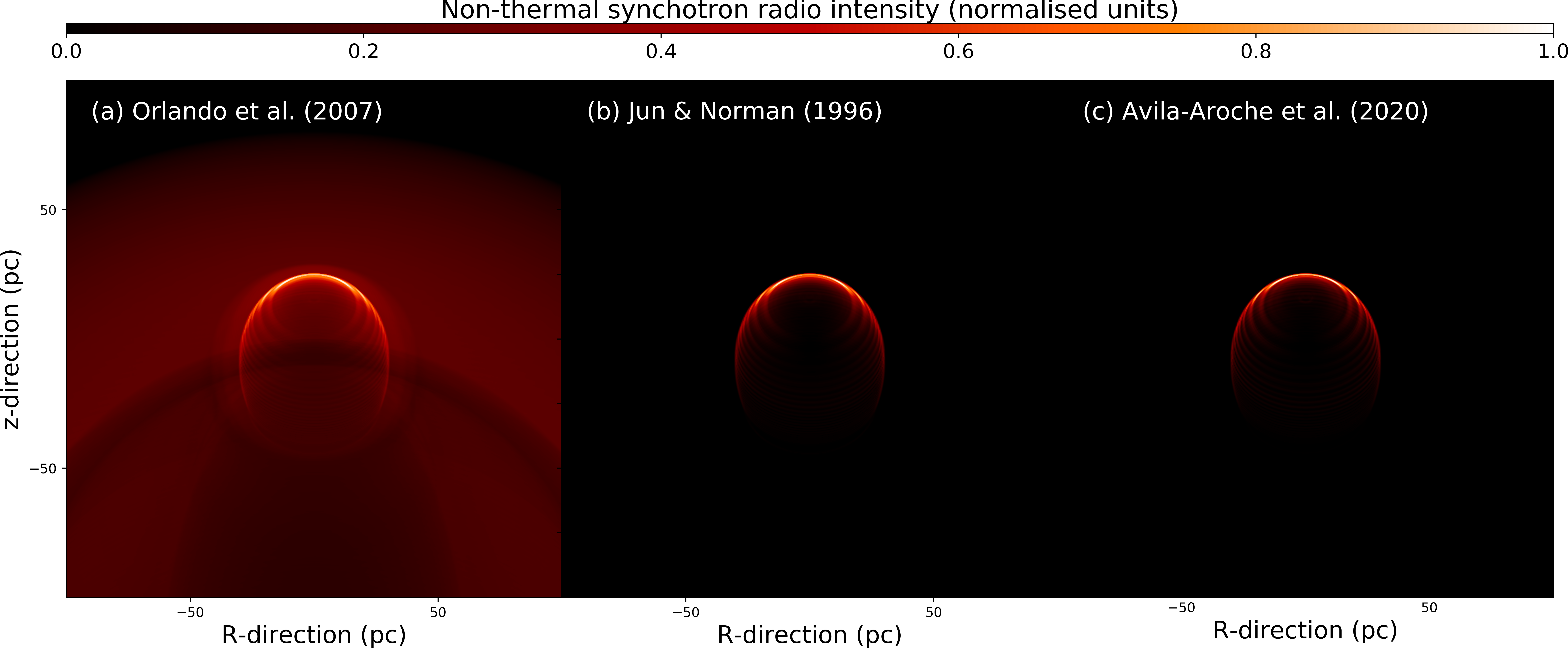}
        \end{minipage}     
        \caption{ 
        \textcolor{black}{
        Normalised maps of radio synchrotron intensity of supernova remnants 
        with progenitor speed $v_{\star}=20\, \rm km\, \rm s^{-1}$ at time 
        $6\, \rm kyr$, and calculated using several prescriptions for the 
        non-thermal emission coefficient. 
        The viewing angle between the equatorial plane and the 
        line of sight is $\theta_{\rm obs}=45\degree$ (middle). 
        }
        }
        \label{fig:radio_map_20kms_app}  
\end{figure*}

\begin{figure*}
        \centering
        \begin{minipage}[b]{ 1.0\textwidth} 
                \includegraphics[width=1.0\textwidth]{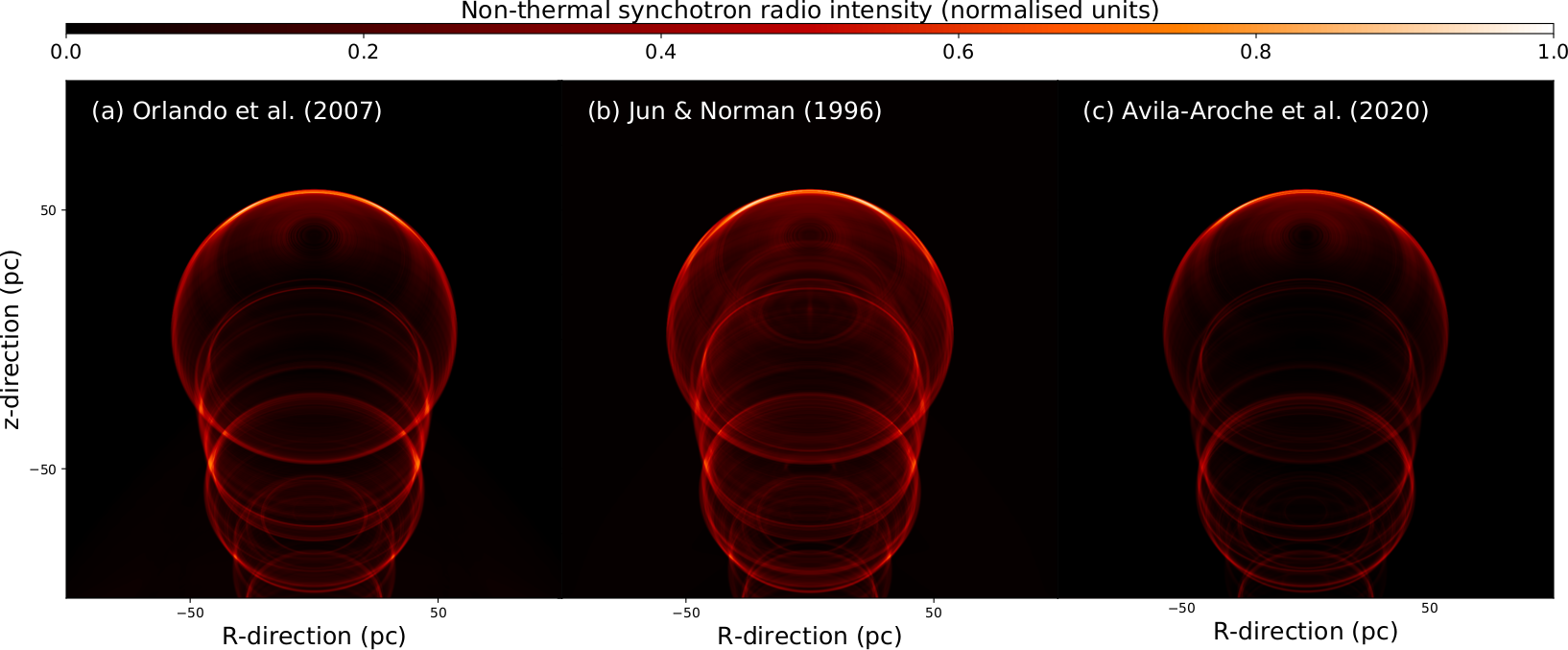}
        \end{minipage}     
        \caption{ 
        \textcolor{black}{
        Same as Fig.~\ref{fig:radio_map_20kms_app} 
        with progenitor speed $v_{\star}=40\, \rm km\, \rm s^{-1}$ at time $80\, \rm kyr$. 
        }
        }
        \label{fig:radio_map_40kms_app}  
\end{figure*}

\section{Emission coefficients for non-thermal synchrotron emission}
\label{sync}

\textcolor{black}{
To the best of our knowledge, three main recipes are available for the emission coefficient 
of non-thermal radio synchrotron emission in the context of supernova 
remnants~\citep{jun_apj_472_1996,orlando_aa_470_2007,avilaaroche_mnras_420_2020}. 
Considering the electron spectrum in the vicinity of the shocks, 
\begin{equation}
        N(E) = K E^{-s},
        \label{eq:N}  
\end{equation}
where $E$ is the electron energy and $s=2 $the index and $K\propto n$. 
They read as, 
\begin{equation}
        j_{\rm sync}^{\rm Orlando}(\nu) \propto K B_{\perp}^{ (s+1)/2 } \nu^{ -(s-1)/2 },
        \label{eq:coeff1}  
\end{equation}
with $\theta_{\rm obs}$ the viewing angle of the observer, $B_{\perp}$ the magnetic 
field component perpendicular to the line of sight and $\nu$ the emission frequency,  
\begin{equation}
        j_{\rm sync}^{\rm Jun}(\nu) \propto K^{2-s} p^{s-1} B_{\perp}^{ (s+1)/2 } \nu^{ -(s-1)/2 },
        \label{eq:coeff2}  
\end{equation}
where $p$ is the gas thermal pressure, and
\begin{equation}
        j_{\rm sync}^{\rm Avila}(\nu) \propto K \textcolor{black}{v'^{2(s-1)}} B_{\perp}^{ (s+1)/2 } \nu^{ -(s-1)/2 },
        \label{eq:coeff3}  
\end{equation}
where $v'$ is the gas velocity in the rest frame of the explosion, respectively. 
This diagnostics has been widely used in, e.g. the context of the core-collapse but also 
type Ia progenitors such as the historical supernova remnants Tycho~\citep{moranchel_mnras_494_2020} 
and SN 1006~\citep{schneiter_mnras_449_2015,velazquez_mnras_466_2017}. 
}

\textcolor{black}{
We generate comparative normalised non-thermal radio emission maps from two selected models of 
supernova remnants. First, one with a progenitor star moving rather slowly with velocity 
$20\, \rm km\, \rm s^{-1}$, and in which the thermal pressure compares with the ram and 
magnetic pressures (Fig.~\ref{fig:radio_map_20kms_app}). Secondly, a model with a fast-moving 
progenitor is moving with velocity $40\, \rm km\, \rm s^{-1}$ and in which the ISM magnetic  
pressure is dynamically unimportant (Fig.~\ref{fig:radio_map_40kms_app}). 
Fig.~\ref{fig:radio_map_20kms_app}a reveals the bright radio synchrotron circumstellar 
medium of the progenitor, produced by wind-ISM interaction before the explosion of the 
massive star, while Fig.~\ref{fig:radio_map_20kms_app}b,c do not. The recipe used in 
Fig.~\ref{fig:radio_map_20kms_app}a clearly overestimates particle acceleration from 
the forward shock of the stellar wind bubble, that is much weaker than the forward 
shock of the expanding supernova blastwave. Hence, the emission coefficient 
in~\citet{orlando_aa_470_2007} is not the most suitable to our core-collapse remnant problem. 
The models calculated with the other emission 
coefficients do not permit to select an optimal one for our 
study (Fig.~\ref{fig:radio_map_20kms_app}b,c).  
The emission coefficient in Eq.~\ref{eq:coeff2} of~\citet{jun_apj_472_1996} has a dependence 
on the thermal pressure $p$, implying that it is sensitive to cooling and heating by 
optically-thin radiative cooling processes and therefore traces the fast shocks well. 
Similarly, the recipe of emission coefficient in Eq.~\ref{eq:coeff3} 
of~\citet{avilaaroche_mnras_420_2020} goes as \textcolor{black}{$j_{\rm sync}^{\rm Avila}(\nu) \propto v^{2}$}, 
that imposes a strong dependence of the chosen frame in which we simulate the 
stellar wind bubble and the supernova explosion. 
Consequently, we decide in our study to use the recipe of~\citet{jun_apj_472_1996}. 
}

%%%%%%%%%%%%%%%%%%%%%%%%%%%%%%%%%%%%%%%%%%%%%%%%%%%%%%%%%%%%%%%%%%%%%%%%%%%%%%%%%%%%%%%%%%%
%%%%%%%%%%%%%%%%%%%%%%%%%%%%%%%%%%%%%%%%%%%%%%%%%%%%%%%%%%%%%%%%%%%%%%%%%%%%%%%%%%%%%%%%%%%
%%%%%%%%%%%%%%%%%%%%%%%%%%%%%%%%%%%%%%%%%%%%%%%%%%%%%%%%%%%%%%%%%%%%%%%%%%%%%%%%%%%%%%%%%%%

\end{document}